\documentclass[12pt]{article}
\newlength{\pubnumber} 

\catcode`\@=11
\@addtoreset{equation}{section}

\def\section{\@startsection{section}{1}{\z@}{3.5ex plus 1ex minus .2ex}
 {2.3ex plus .2ex}{\large\bf}}
\def\subsection{\@startsection{subsection}{2}{\z@}{2.3ex plus .2ex}
 {2.3ex plus .2ex}{\bf}}
\newcommand\Appendix[1]{\def\thesection{Appendix \Alph{section}}
 \section{\label{#1}}\def\thesection{\Alph{section}}}

 \usepackage{graphicx}
\begin{document}

\begin{titlepage}
\samepage{
\setcounter{page}{0}
\rightline{\tt hep-th/0610319}
\rightline{October 2006}
\vfill
\begin{center}
    {\Large \bf  
  Fighting the Floating Correlations:\\
Expectations and Complications in Extracting Statistical 
Correlations from the String Theory Landscape\\}
\vfill
\vspace{.10in}
   {\large
      Keith R. Dienes\footnote{
     E-mail address:  dienes@physics.arizona.edu},$\,$ 
    Michael Lennek\footnote{E-mail address:  mlennek@physics.arizona.edu}}
\vspace{.10in}

 {\it  Department of Physics, University of Arizona, Tucson, AZ  85721  USA\\}
\end{center}
\vfill
\begin{abstract}
  {\rm 
   The realization that string theory gives rise to a huge landscape
   of vacuum solutions has recently prompted a statistical approach 
   towards extracting phenomenological predictions from string theory.
   Unfortunately, for most classes of string models, direct enumeration 
   of all solutions is not computationally feasible and thus
    statistical studies must resort to other methods in order to 
    extract meaningful information.
   In this paper, we discuss some of the issues that arise 
   when attempting to extract statistical correlations from a 
   large data set to which our computational access is necessarily limited.
   Our main focus is the problem of ``floating correlations''.
    As we discuss, this problem is  
   endemic to investigations of this type and
     reflects the fact that not all physically distinct 
    string models are equally likely to be sampled in any random   
    search through the landscape, thereby causing statistical correlations
    to ``float'' as a function of sample size.
   We propose several possible methods that can be used 
   to overcome this problem, and we show
    through explicit examples that these methods  
    lead to correlations and statistical distributions 
    which are not only stable as a function of sample size, 
         but which differ significantly from those which 
    would have been na\"\i vely apparent from only a partial data set.  }
\end{abstract}
\vfill
\smallskip}
\end{titlepage}

\setcounter{footnote}{0}

\def\beq{\begin{equation}}
\def\eeq{\end{equation}}
\def\beqn{\begin{eqnarray}}
\def\eeqn{\end{eqnarray}}
\def\half{{\textstyle{1\over 2}}}

\def\calO{{\cal O}}
\def\calE{{\cal E}}
\def\calT{{\cal T}}
\def\calM{{\cal M}}
\def\calF{{\cal F}}
\def\calY{{\cal Y}}
\def\calV{{\cal V}}
\def\calN{{\cal N}}
\def\ibar{{\overline{\i}}}
\def\qbar{{\overline{q}}}
\def\mm{{\tilde m}}
\def\ahat{{\hat a}}
\def\nn{{\tilde n}}
\def\rep#1{{\bf {#1}}}
\def\ie{{\it i.e.}\/}
\def\eg{{\it e.g.}\/}

\def\boxit#1{\vbox{\hrule\hbox{\vrule\kern3pt
\vbox{\kern3pt#1\kern3pt}\kern3pt\vrule}\hrule}}

\def\Str{{{\rm Str}\,}}
\def\bone{{\bf 1}}

\def\thetai{{\vartheta_i}}
\def\thetaj{{\vartheta_j}}
\def\thetak{{\vartheta_k}}
\def\thetaibar{\overline{\vartheta_i}}
\def\thetajbar{\overline{\vartheta_j}}
\def\thetakbar{\overline{\vartheta_k}}
\def\etainv{{\overline{\eta}}}

\def\modinvmeasure{{  {{{\rm d}^2\tau}\over{\tautwo^2} }}}
\def\qbar{{  \overline{q} }}
\def\ahat{{ \hat a }}

\newcommand{\newc}{\newcommand}
\newc{\gsim}{\lower.7ex\hbox{$\;\stackrel{\textstyle>}{\sim}\;$}}
\newc{\lsim}{\lower.7ex\hbox{$\;\stackrel{\textstyle<}{\sim}\;$}}

\hyphenation{su-per-sym-met-ric non-su-per-sym-met-ric}
\hyphenation{space-time-super-sym-met-ric}
\hyphenation{mod-u-lar mod-u-lar--in-var-i-ant}


\def\inbar{\,\vrule height1.5ex width.4pt depth0pt}

\def\IC{\relax\hbox{$\inbar\kern-.3em{\rm C}$}}
\def\IQ{\relax\hbox{$\inbar\kern-.3em{\rm Q}$}}
\def\IR{\relax{\rm I\kern-.18em R}}
 \font\cmss=cmss10 \font\cmsss=cmss10 at 7pt
\def\IZ{\relax\ifmmode\mathchoice
 {\hbox{\cmss Z\kern-.4em Z}}{\hbox{\cmss Z\kern-.4em Z}}
 {\lower.9pt\hbox{\cmsss Z\kern-.4em Z}}
 {\lower1.2pt\hbox{\cmsss Z\kern-.4em Z}}\else{\cmss Z\kern-.4em Z}\fi}

\long\def\@caption#1[#2]#3{\par\addcontentsline{\csname
  ext@#1\endcsname}{#1}{\protect\numberline{\csname
  the#1\endcsname}{\ignorespaces #2}}\begingroup
    \small
    \@parboxrestore
    \@makecaption{\csname fnum@#1\endcsname}{\ignorespaces #3}\par
  \endgroup}
\catcode`@=12

\input epsf

\section{Introduction}
\setcounter{footnote}{0}
\label{intro}
\setcounter{footnote}{0}

Over the past few years,
it has become increasingly clear that string theory gives rise 
to a very large number of vacuum solutions~\cite{landscape}.  
Because the specific
low-energy phenomenology that can be expected to emerge from the string depends
critically on the particular choice of vacuum state,
detailed quantities such as particle masses
and mixings --- and even more general quantities and structures such as the choice of gauge
group, number of chiral particle generations, magnitude of the supersymmetry-breaking scale,
and the cosmological constant --- can be expected to vary significantly from one vacuum
solution to the next.
Thus, in the absence of some sort of vacuum selection principle,
it has been proposed that
meaningful phenomenological predictions from string theory might instead be
extracted statistically, through the discovery of statistical correlations 
across the huge ``landscape'' of string vacua~\cite{abstract}. 
Such string-derived correlations would relate
different phenomenological features that are apparently unrelated in field theory,
and would thus represent string-theoretic predictions that hold for the majority of
string vacua. 

Unfortunately, the space of possible vacua is extremely large, with 
some estimates putting the number of phenomenologically interesting vacua at 
$10^{500}$ or more~\cite{abstract}.
Direct computational access to this large data set is therefore virtually impossible,
and one is forced to undertake statistical studies of a more limited nature.

To date, there has been considerable work in this 
direction~\cite{abstract,abstract2,susybreakingabstract,NPcomplete,direct,direct2,direct3,dienes};
for reviews, see Refs.~\cite{review1,review2}.
Collectively, this work has focused on different classes of string models,
both closed and open, employing a number of different underlying string
constructions and formulations.
However, regardless of the particular string model or construction procedure
utilized, any such statistical analysis can be characterized as belonging
to one of three different classes:
\begin{itemize}
\item  {\bf Abstract studies:}  First, there are abstract mathematical studies  
       that proceed directly from
       the construction formalisms (\eg, considerations of flux
       combinations).  Although large sets of specific string models
       are not enumerated or analyzed, general expectations and trends 
       are deduced based
       on the statistical properties of the parameters that are relevant
       in these constructions. 
\item  {\bf Direct enumeration studies:}  Second, there are statistical studies based on direct 
       enumeration of finite subclasses of string models.
       Within these well-defined subclasses, one enumerates
       literally all possible solutions and thereby collects statistics 
       across a large but finite tractable data set.
\item  {\bf Random search studies:}  Finally, there are statistical studies that aim
       to explore a data set which is (either effectively or literally) infinite
       in size.  Such studies involve randomly 
       generating a large but finite sample of actual string models and
       then analyzing the statistical properties of the sample, assuming the
       sample to be representative of the class of models under examination as a whole.
\end{itemize}
Indeed, all three types of studies have been undertaken in the literature.

Certain difficulties are inherent to all of these approaches.
For example, in each case there is the over-arching problem
of defining a measure in the space of string solutions.
We shall discuss this problem briefly below, but this is 
not the chief concern of the present paper and for simplicity
we shall simply assume that each physically distinct
string model is to be weighted equally in any averaging process. 

By contrast, other difficulties are specifically tied to 
individual approaches.  For example, the first approach has great 
mathematical generality
but often lacks the precision and power that can come from direct 
enumerations of actual string models.
Likewise, the second approach is fundamentally limited
to classes of string models for which a full enumeration is possible ---
\ie, string constructions which admit a number of solutions
which is both finite and accessible with current computational power. 

For these reasons, the third approach might ultimately seem to have 
the best prospects for generating precise statistical statements
about a relatively large string landscape.  
As has been recently shown, the problem of directly enumerating 
certain classes of string models is actually NP-complete~\cite{NPcomplete}.  
This fact implies that our computational access to the string 
landscape will always be quite limited, which in turn suggests 
that random search studies may be more efficient for exploring the string landscape.  
Indeed, most large-scale census studies are of this type.

Although significant effort has been devoted to studying 
the algorithms and issues facing direct enumeration studies,
relatively little effort has been invested 
in studying the issues facing random search studies.  
In this paper, we shall present some elementary observations concerning 
some of the potential pitfalls of such studies, and the methods by which they
can be overcome. 

Clearly, one fundamental difficulty is that one must assume that  
the sample set of string vacua is representative of the 
relevant class of string vacua as a whole.  
To attempt to ensure this, one typically generates these sample 
sets as randomly as possible from amongst the functionally infinite set of 
allowed solutions in the class. 
One therefore assumes that no bias has been introduced into
this procedure.
However, as we shall discuss in this paper, 
there is a unique alternative kind of bias which is nearly inevitable 
in random searches through the string landscape. 
Moreover, as we shall explain, this bias leads directly to the problem of 
``floating correlations''.
This in turn leads to tremendous distortions in the statistical
correlations that one would appear to extract through such studies. 

In this paper, we shall begin by discussing the origins of this phenomenon.
We shall then discuss various means by which it may be overcome.
Finally, we shall present an explicit example drawn from studies of the
heterotic landscape which illustrates that these issues, and their resolutions,
can dramatically alter the magnitudes of the correlations that 
one would na\"\i vely appear to extract from the landscape.

\section{The problem of floating correlations}
\setcounter{footnote}{0}
    
In general, there are many different construction formalisms which may be employed
in order to build self-consistent string models.  For example, closed string models may be constructed
through orbifold techniques (with or without Wilson lines), or alternatively
using geometric techniques (\eg, by specifying particular Calabi-Yau compactifications).
There are also generalized conformal field theory techniques (such as those utilized
in Gepner constructions), or special cases of these which 
involving only free worldsheet bosons or fermions with different boundary-condition
phases.
Similar choices exist as well for open strings, where one can have, \eg, intersecting
D-branes, fluxes, and so forth~\cite{constructions}.
Not all construction formalisms are distinct, and the sets of models 
which can be realized through each construction technique
can often have significant overlaps.

Within each construction formalism, there are certain free parameters 
which one is free to choose;
we shall collectively label these internal parameters $\lbrace x_i\rbrace$.
These may be compactification moduli, boundary-condition phases, Wilson-line coefficients, 
or topological quantities specifying Calabi-Yau manifolds;
likewise they might be D-brane dimensionalities and charges, 
wrapping numbers, or intersection angles.
We may also include among the set $\lbrace x_i\rbrace$ the vevs of moduli fields
and/or fluxes
which are necessary for guaranteeing stable (or at least sufficiently flat) vacuum solutions.
As long as these internal parameters are chosen to satisfy whatever self-consistency constraints
are inherent to the relevant construction method (such as those stemming from conformal invariance
and modular invariance in the case of closed strings, or anomaly and tadpole cancellations 
in the case of open strings), one is guaranteed to have constructed a 
bona-fide string model.  
 
However, regardless of the particular construction formalism employed, one cannot 
generally define a given string model as being distinct from all others on the basis 
of an examination of these parameters $\lbrace x_i\rbrace$.
Rather, one must deduce the spacetime properties of the resulting string model 
in order to deduce whether this model is truly unique when compared with another.     
Such spacetime properties might include, for example, 
the gauge group, the number of spacetime supersymmetries,  
the entire particle spectrum, and the associated couplings.
Collectively, we can describe these spacetime properties as belonging to a set
of spacetime parameters $\lbrace y_j\rbrace$. 
If any of these $y$-parameters are different for two candidate models, we 
say that the two candidate models are truly distinct --- \ie, that
we truly have two models. 
Of course, the parameters $\lbrace y_j\rbrace$ are not independent of
each other (as they might be in field theory), 
but are presumably correlated by the fact that they emerge
from a given self-consistent string model.
These are the types of correlations that one ultimately hopes to 
extract as string predictions from the landscape.  

In general, each construction technique provides a recipe or
prescription for starting with a self-consistent set of parameters
$\lbrace x_i\rbrace$ and generating a corresponding set of spacetime
parameters $\lbrace y_j\rbrace$.  
In other words, each construction formalism implicitly provides 
us with a set of functions $f_j$ such that
\beq
                     y_j ~=~ f_j(\lbrace x_i\rbrace )~.
\eeq
However, deriving the exact explicit form of such functions is a formidable task,
and it is not always possible to extract these functions explicitly from the underlying
construction method.
What is important for our purposes, however, is that such functions represent the dependence
of the spacetime $y$-parameters on the internal 
$x$-parameters.
 
\begin{figure}[htb]
\centerline{
   \epsfxsize 4.0 truein \epsfbox {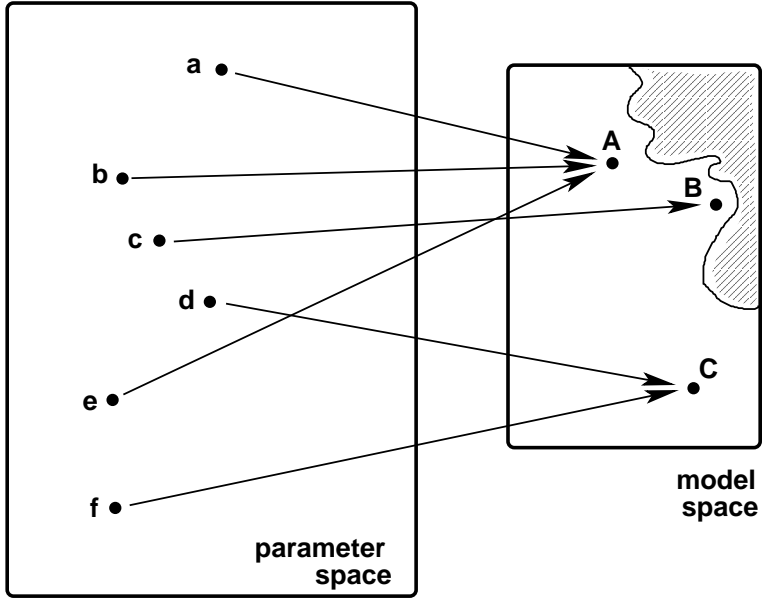}
    }
\caption{Each string model-construction technique provides a mapping between
a space of internal parameters (such as compactification moduli) and a physical
model in spacetime.  However, this mapping is not one-to-one, and there
generally exists a huge redundancy wherein a single physical string model 
in spacetime (such as Model~A in the figure) can have multiple redundant realizations
in terms of internal parameters.  For this reason, the space of internal parameters
is usually significantly larger than the space of obtainable distinct models.
The shaded region represents models which, though entirely self-consistent, are
not realizable through the construction technique under study.}
\label{figone} 
\end{figure}

Although not much is generally known about such functions $f_j$, one thing is clear:
these functions are not one-to-one.
Rather, there exist numerous {\it redundancies}\/ according to which different combinations
of $\lbrace x_i\rbrace$ can lead to exactly the same $\lbrace y_j\rbrace$. 
In general, such redundancies exist because of a variety of factors.
Sometimes, there are underlying identifiable worldsheet symmetries (often
of a geometric nature, \eg, mirror symmetries) which
cause two different constructions to lead to the same physical string model.
In such cases, these redundancies are well-understood and can perhaps be quantified
and eliminated from the model-construction procedure,
but this process becomes extremely intractible and inefficient for sufficiently
complicated models.
In other cases, however, there may simply be redundancies in the 
chosen construction formalism
such that different combinations of parameters can result in the same physical
string model in spacetime.
For example, it often happens that two unrelated sets of
orbifold twists and Wilson lines can result in the same string model
even when there is no apparent geometric connection between them.
Regardless of the cause, however, the important point is that the mapping 
between the internal $x$-parameters and the spacetime $y$-parameters is 
not one-to-one.
We therefore are faced with the situation sketched in Fig.~\ref{figone}.

\begin{figure}[htb]
\centerline{
   \epsfxsize 4.0 truein \epsfbox {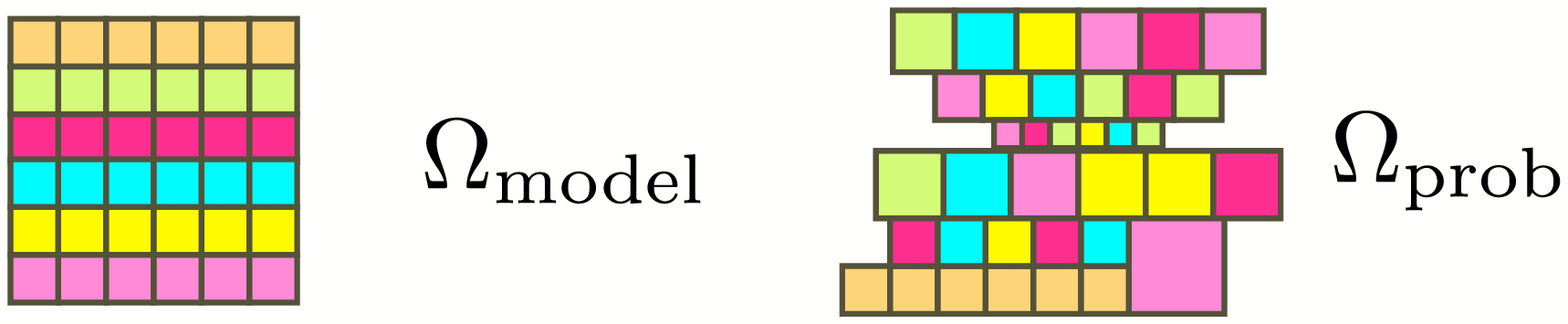}
    }
\caption{Illustration of the difference between the model space and
    the probability space, with total volumes $\Omega_{\rm model}$ and
    $\Omega_{\rm prob}$ respectively.  
   Each box represents a distinct string model.  In the model space, each model occupies the same 
   volume, whereas in the probability space, each model occupies a volume 
    which is proportional to its probability 
   of production.  Note that the shaded regions 
          of Fig.~\protect\ref{figone} have zero probability 
    of being produced and thus do not appear in the probability space at all.}
\label{ModSpace} 
\end{figure}

This feature can have devastating consequences for a random search through
the space of string models.
Because any such search must be tied to a particular construction technique,
one cannot simply survey the model space of self-consistent $\lbrace y_j\rbrace$;  
rather, one is forced to survey the parameter space $\lbrace x_i\rbrace$.
This means that we do not have direct access to the model space in which each model is weighted
equally;  rather, we only have access to
deformation of this model space in which models with multiple $x$-representations
occupy a larger effective volume than those with fewer $x$-representations.
We may refer to this deformed model space as a {\it probability space}\/,  
since each model in the probability space shall be defined to occupy a volume 
which is proportional to its probability of being selected through a 
generation of self-consistent
$x$-parameters.
This is illustrated in Fig.~\ref{ModSpace}.

This can lead to three potential types of bias in a random model search.
The first two are relatively obvious and straightforward to deal with:
\begin{itemize} 
\item  First, one may not be sampling the parameter space in a truly random way.
        Indeed, the selection of
        $x$-parameters may be skewed as the result of 
         a systematic algorithmic or computational bias.
      However, this kind of bias is not the focus of this paper, and we shall assume
       that our computational algorithms provide a truly random sampling of 
     model-construction parameters.
     (In any case, the methods we shall eventually be developing in this
     paper can compensate for a bias of this type as well.)
\item Second, one might be oversampling models for which there exist
          multiple internal realizations.  For this reason, it is necessary
         to ensure that one does not consider a given string model more   
         than once in the random search process.  In other words, 
          each time a self-consistent set of $x$-parameters is
          generated, one must calculate the corresponding $y$-parameters
         {\it and verify that these parameters do not match those 
         of any other model which has previously been considered in the 
          same sample}.  While conceptually straightforward, this requirement
         is computationally and memory intensive since it requires that any
           search procedure maintain a cumulative, readable memory of all models that have            
          already been constructed in the sample. 
            Indeed, we have found that this feature alone tends to provide 
          the most severe limitations on the sizes of string model samples
         that can feasibly be generated.
\end{itemize}
Thus, while these types of bias are important, both can easily
be addressed.  

However, the third type of bias is more subtle and is
the focus of this paper.
In some sense, this problem is the reverse of the second problem itemized
above:  some models are relatively {\it hard}\/ to generate in terms
of appropriate $\lbrace x_i\rbrace$.
Of course, this would not be an issue if the redundancy indicated in Fig.~\ref{figone}
were relatively evenly distributed across the model space.
Counting each model with a multiple redundancy only once
would then eliminate all bias.
However, it turns out that {\it some models have redundancies which are 
greater than those of other models by many, many orders of magnitude}.
What this means in practice is that when one is randomly sampling  
the parameter space, one easily ``discovers'' models such as Model~A
in Fig.~\ref{figone} while never finding models such as Model~B. ~Thus, 
while models such as Model~A are likely to be included 
in any random sample of string models,
models such as Model~B are almost certain to be missed.
Indeed, in a typical search, we are not likely to have the computational
power to probe even the full set of highly likely models.
Thus we are almost certain to under-represent the relatively unlikely models,
assuming we find such models at all.   

This kind of disparity is of little consequence if all physical properties of
interest are evenly distributed across the model space.
For example, if we are interested in knowing what fraction of models have chiral
spectra, this kind of disparity will be irrelevant if the chirality 
property is uncorrelated with the redundancy property. 
However, it is usually the case that the very same underlying features which create
the hierarchy of redundancies for different string models
also lead to uneven distributions with respect
to their physical properties.  
For example, a given string construction method may easily yield a set of
models with a given property (\eg, a gauge group of a given large, fixed rank), and yet
be capable of yielding models that do not have that property (\eg, which exhibit rank-cutting)
in some carefully fine-tuned circumstances. 
Thus, if we are generating a sample set of models, 
we are likely to miss those ``rare'' models until our sample size becomes extremely large.

The implications of this can be rather severe.
  {\it If the physical property about which we are seeking statistical information  
happens to correlate with this redundancy, then our statistical correlations
or percentages will necessarily evolve (or ``float'') as a function of the sample size.}
Even worse, because we cannot hope to approach a complete saturation of the 
model space and because we have little guidance as to the sizes or patterns of
redundancy in our model-construction procedures, we cannot obtain meaningful statistics 
by generating more models and waiting for this floating process to become stable.
We emphasize that this is a problem that must be faced {\it regardless}\/ of our
choice of model-construction technique and {\it regardless}\/ of how carefully we construct
randomized algorithms for model generation.

Thus, we may summarize this problem as follows.
Each time we construct a self-consistent set of $x$-variables $\lbrace x_i\rbrace$,
we examine the corresponding $y$-variables to see if we have really constructed
a new model that we have not seen before.
If so, we add it to our sample set of models;  if not, 
we disregard the set $\lbrace x_i\rbrace$ and generate another.
Very soon, we reach a stage at which models with some physical properties are
``common'', and models with other physical properties are ``rare''.
However, it is {\it a priori}\/ impossible to determine what percentages 
of models might be ``common'' and what percentages are ``rare'' on the basis of
this sample set.
The problem is that if we keep generating new candidate 
sets $\lbrace x_i\rbrace$,
we will tend not to generate any further models of the ``common'' variety
because they will have already been generated.
In other words, each additional distinct model that we generate has an increasing
probably of being ``rare'', which is why it is distinct from those that have already
been constructed.
Thus, rare properties tend to become less rare as the sample size increases,
which causes our statistical correlations to float as functions of the sample size. 
Indeed, in most realistic situations, this problem can be further compounded
by the fact that physically interesting properties such as 
spacetime SUSY, gauge groups, numbers of chiral generations, 
and so forth may be differently distributed across models with varying 
intrinsic probabilities of being selected.
This too causes our statistical correlations to float as functions of the sample size.

This, then, is the problem of floating correlations.
What is required is a means of overcoming this type of bias and extracting
statistical information, however limited, from such a model search.

\section{Modelling the model search:  Drawing balls from an urn}
\setcounter{footnote}{0}
\label{Balls}

It will help to develop
a mathematical model for the process of randomly exploring the model space.  
Towards this end,
let us begin by imagining a big urn filled with balls of different colors
and compositions.
For example, some of the balls are red, while others are blue;
likewise, some of the balls are plastic, while others are rubber.
Each ball shall correspond to a distinct string model.
Thus, exploring the string model space through the random generation of string models
becomes analogous to the act of drawing a ball from the urn, noting
its properties, marking it for future identification, replacing the ball in the urn,
mixing, and then repeating over and over.
Of course, since we replace each ball after we have drawn it, each draw is independent.

If the model-generation method is truly random without any inherent biases,
then each ball will have the same basic probability to be 
drawn regardless of its properties.  We shall examine this case in detail 
first, and then consider more realistic cases where the model-generation method is biased.

Clearly, each draw from the urn need not result in a new ball because there
is a possibility that we will draw 
a ball that has already been seen. 
Thus, after $D$ draws from the urn, we will have found a number of models
$M(D)$ which we expect to be smaller than $D$.
Although $M(D)$ is restricted to be an integer, it is straightforward to derive
an expression for the expectation value $\langle M(D)\rangle$.
If we have a total of $N$ different balls in the urn (so $N$ distinct models 
are realizable by our construction method), then the probability of drawing 
any specific ball is simply $P = 1/N$.
Since we are exploring the model space randomly, 
the difficulty of finding a {\it new}\/ ball will be related to how fully
explored the model space already is.  
If we have already seen $x$ distinct  balls, then
the probability that a new draw will yield a previously unseen ball is given
by 
\beq
               P_{\rm new} ~=~ 1-\frac{x}{N}~.
\label{newmodelpop}
\eeq
Given this, we can determine $\langle M(D)\rangle$ recursively.
If $\langle M(D)\rangle$ is already known, then clearly
\beqn
                  \langle M(D+1) \rangle &=&  
                    \left[ 1-{M(D)\over N}\right] \,\left[ \langle M(D)\rangle +1\right] ~+~ 
                     \left[ {M(D)\over N}\right] \,\langle M(D)\rangle ~\nonumber\\
                      &=& \langle M(D)\rangle\,\left[ 1-{1\over N}\right] ~+~ 1~.
\label{recursion}
\eeqn
Here the first term on the first line reflects the contribution from the possibility that the 
next draw yields a new ball hitherto unseen, while the second term reflects the
possibility that it does not;  moreover, in passing to the second line we have replaced 
$M(D)$ by $\langle M(D)\rangle$.  
This recursion relation, along with the initial condition $\langle M(0)\rangle = M(0)=0$, allows us to
solve for $\langle M(D)\rangle$ exactly:
\beq   
                  \langle M(D) \rangle ~=~  \sum_{j=0}^{D-1} \left( 1-{1\over N}\right)^j~=~ N\left\lbrack
              1 -\left( 1-{1\over N}\right)^D\right\rbrack~,
\label{Onepop}
\eeq
and for $N\gg 1$ we may approximate this as
\beq
               \langle M(D)\rangle ~\approx~ N(1-e^{-D/N})~.
\label{Onepopulationmodel}
\eeq

This has the basic behavior we expect;  indeed, calculations along these lines have appeared
more than a decade ago in Ref.~\cite{Senechal}.
When the model space is relatively 
unexplored, it is not difficult to find a new model, but as the model space 
becomes more explored it gets 
harder to find new models.  
The main feature to note here is that the 
total number $N$ of distinct models appears both as a 
multiplicative factor and in the exponent in this expression.
This only happens when all models have an equal probability to be generated.

Unfortunately, as we have discussed in Sect.~2, 
it will typically be the case that different models will have different
probabilities of being generated.
Indeed, as discussed in Sect.~2, what we are exploring randomly 
is typically not the
model space, but rather the probability space.

In order to account for this, let us now modify the above analysis by imagining that 
each ball in the urn has a different intrinsic probability of being drawn from
the urn, and that this probability
depends on its composition.  
For example, we may imagine that plastic balls are intrinsically
smaller or lighter than rubber balls, and thus have a smaller 
cross section for being selected when
we reach into the urn. 
In general, we shall let $p_i$ denote the relative intrinsic probability 
that a ball in population $i$ will be selected on a random draw,
and we shall let $N_i$ denote the sizes of these populations.
For example, if $N_{\rm plastic}/N_{\rm rubber}=1/3$
but $p_{\rm plastic}/p_{\rm rubber}=1/4$, then 
a plastic ball will be $1/12$ as likely to be drawn from the urn
as a rubber ball.
We shall assume that all of the balls with a common composition $i$ share a common
intrinsic probability $p_i$ of being selected, but we shall make no assumption
about the number of such populations.
Also note that only ratios of the different $p_i$ shall matter, so there is no need
to normalize the $p_i$ in any particular fashion.\footnote{
        We emphasize that in the case of actual string model-building,
          models do not have an intrinsic $p_i$ except in the presence  
          of a particular model-generation technique.  
          Thus, the $p_i$'s are associated not only with a given set of 
           models, but also with a specific model-generation technique.
           As a practical matter, however,
           one must always have a formalism through which to generate
           models, so it is sufficient to associate the $p_i$'s with the 
           models themselves, as we have done with this ball/urn analogy.
            We also note that even if there is a bias {\it within}\/  
           the model-generation technique, so that the parameter space
            in Fig.~\ref{figone} is not explored truly randomly, this effect can also
             be incorporated within the probabilities $p_i$ so long as 
             each parameter combination is explored at least once.
            Thus, the methods that we shall be developing for overcoming 
              production biases can overcome this type of bias as well.}
           
As illustrated in Fig.~\ref{ModSpace},
each model occupies an equal volume in the model space but only a rescaled
volume in the probability space;  the probabilities $p_i$ describe these
rescalings. 
Indeed, the total volumes of the model and probability spaces 
can be defined as
\beq
            \Omega_{\rm model} ~=~ \sum_{i} N_i~, 
      ~~~~~\Omega_{\rm prob}  ~=~ \sum_{i}   p_i N_i~. 
\label{probspace}
\eeq
Of course, 
with this definition
$\Omega_{\rm prob}$ will scale with the overall normalization of the 
$p_i$, but this will not be
relevant in  the following.
What is important, however, is that the probability space will 
be different from the model space  
if all of the $p_i$ are not identical. 
Thus, the volume relations amongst populations with different 
$p_i$ will be different in the two spaces.  However, by construction, the 
volume relations amongst models with 
the {\it same}\/ $p_i$  will be the same in both the model and probability spaces.  

We are now in a position to address what will happen as this model space is explored.  
By definition, the probability of drawing a ball from a given population is directly
related to the volume occupied in the probability space
by that particular population.  
For a given population $i$, this probability is $\hat P_{i} = p_i N_i/ \Omega_{\rm prob}$.  
However, the probability of finding a {\it new, previously unseen}\/ ball within the population $i$ 
will also depend on the number 
$x_i$ of distinct balls from population $i$ which have already been found:
\beq
             \hat P^{(i)}_{\rm new}~=~   \frac{p_i}{\Omega_{\rm prob}}~(N_i - x_i)~.
\label{probinewmodel}
\eeq
 Here the notation $\hat P$ (rather than $P$) indicates 
that the probability $\hat P$ in Eq.~(\ref{probinewmodel}) is entirely {\it unrestricted}\/, \ie,
there is no prior assumption that the draw will even select a model from the $i$-population.
Using this equation, we can follow our previous steps to calculate the expected number 
of distinct $i$-models $\langle M_i(D)\rangle$ that will be found 
after $D$ draws from the urn.
Our recursion relation takes the form
\beq
          \langle M_i(D+1) \rangle ~=~ \langle M_i(D) \rangle 
             \left[ 1- {p_i\over \Omega_{\rm prob}}\right] ~+~ {p_i N_i\over \Omega_{\rm prob}}~,
\eeq
and with the initial condition $\langle M_i(0)\rangle =M_i(0)=0$ we find the solution 
\beq
           \langle M_i(D)\rangle ~=~ N_i 
      \left[  1- \left( 1- {p_i\over \Omega_{\rm prob}} \right)^D\right]~\approx~ 
             N_i \left( 1- e^{-D p_i/\Omega_{\rm prob}} \right)~.
\label{otherprobinewmodel}  
\eeq
Note that the prefactor no longer matches the factor in the exponential. 

Eqs.~(\ref{probinewmodel}) and (\ref{otherprobinewmodel}) give us a general sense of when different 
populations of models will be found.  The populations with the largest $p_i$ will begin to be 
explored first simply because they have the larger probabilities of being selected.
This will then increase $x_i$, which in turn 
makes it more difficult to find new models in this population.
Subsequent new models that are found will then start to preferentially come from populations
with $p_j<p_i$.  
Indeed, only when
\beq
               p_j ~=~ \left(\frac{N_i-x_i}{N_j-x_j}\right)  \, p_i
\label{goldenrat}
\eeq
will the probability of drawing a new model from the model space of population $j$ be equal 
to that of drawing a new model from the model space of population $i$.  
Thus, the exploration of model spaces with smaller $p_i$ 
will always lag the exploration of spaces of models with larger $p_i$.

Note that Eq.~(\ref{otherprobinewmodel}) describes the growth of the individual
quantities $\langle M_i(D)\rangle$ as functions of the number $D$ of draws.
For this purpose, any selection from the urn counts as a draw.
However, for some purposes, it is also useful to define the {\it restricted}\/ draw
$d_i$ which denotes the number of times a ball from population $i$ is drawn
(again regardless of whether this ball has previously been seen).
Each time $D$ increases by one, we can be certain that one and only one of the $d_i$ 
increases by one because our probability populations are disjoint. 
However, the {\it expectation value}\/ 
$\langle d_i\rangle$ will be given by
\beq
            \langle d_i\rangle ~=~  {p_i N_i\over \Omega_{\rm prob}}\, D~.
\eeq
Using $\langle d_i\rangle$, we can therefore 
rewrite Eq.~(\ref{probinewmodel}) in the form
\beq
           \langle M_i(D)\rangle ~=~ N_i \left[  1- \left( 1- {1\over N_i}
                \right)^{\langle d_i\rangle} \right]~\approx~ 
             N_i \left( 1- e^{-\langle d_i\rangle /N_i} \right)~.
\label{probinewmodel2}  
\eeq
Of course, as expected, these results have the same forms as
Eqs.~(\ref{Onepop}) and (\ref{Onepopulationmodel}) since the use of the
restricted draw $\langle d_i\rangle$ allows us to consider each population
as truly separate in the drawing process.

Even at this stage, we have still not completely modelled  
the string model-exploration process.
This is because we cannot assume that the physical characteristics 
of a given model
(such as its degree of supersymmetry, the rank or content of its gauge group,
the chirality of its spectrum, or its number of fermion generations)
are correlated in any way with its probability of being drawn.
Or, to continue with our analogy of the balls in the urn, 
even though the plastic balls may be smaller or lighter than the rubber
balls (thereby giving the plastic balls a smaller intrinsic probability
$p_i$ of being drawn than the rubber balls), 
the physical characteristics of the string model may correspond
to a completely independent variable such as the {\it color}\/ of the ball.
Some balls may be red 
and some balls may be blue,
and we have no reason to assume that all red balls are plastic or that
all blue balls are rubber.
In the following, therefore, we shall continue to let the composition
$i$ of the balls 
represent their probabilities
of being selected, but we shall also let
the color $\alpha$ of the ball (red, blue, {\it etc.}) denote its physical characteristics.
This is consistent with the conventions in Fig.~\ref{ModSpace}, where different colors/shadings
denote different physical characteristics while size rescalings denote different probabilities
of being drawn. 

Note that while the different probability populations are necessarily disjoint,
the physical characteristic classes need not be disjoint at all.  For example,
two classes $\alpha$ and $\beta$ may have a partial overlap, such as would
occur if characteristic $\alpha$ denotes the presence of an $SU(3)$ gauge-group factor
while $\beta$ denotes the presence of ${\cal N}=1$ spacetime SUSY;
alternatively, one class may be a subset of another,
as would occur if $\alpha$ denotes the presence of $SU(3)$ while
$\beta$ denotes the presence of the entire Standard-Model gauge group.
All that is required in our formalism is that each class correspond to a
set of models exhibiting a well-defined set of particular physical characteristics.

Given this, the populations will generally 
fill out a {\it population matrix}\/ $N_{\alpha i}$.  
Moreover, given this population matrix, 
it is then straightforward to determine the average expected   
numbers of distinct models with particular sets of physical characteristics: 
\beq
         \langle M_\alpha(D)\rangle ~=~ \sum_i  N_{\alpha i} \left[
            1 - \left( 1- {p_i\over\Omega_{\rm prob}}\right)^D \right]
          ~\approx~ \sum_i N_{\alpha i} \left( 1- e^{-D p_i /\Omega_{\rm prob}} 
            \right)~.
\label{bigeq}
\eeq
Thus, as we draw balls from the urn and note their physical properties,
we expect the numbers of distinct models exhibiting particular physical
properties to grow as a sum of weighted exponentials, where each exponential
is weighted by its own population size $N_{\alpha i}$ and where
the ``time constant'' for each exponential is related to a unique
probability fraction $p_i$.

Clearly, each different physical characteristic $\alpha$ can be expected 
to have its own unique pattern of growth for $\langle M_\alpha(D)\rangle$ as a 
function of $D$.
However, it may occasionally happen that two
different physical characteristics $\alpha$ and $\beta$ will
nevertheless give rise to quantities 
$\langle M_\alpha(D)\rangle$ and
$\langle M_\beta(D)\rangle$ which share the same overall behavior as functions
of their arguments, with at most only an overall rescaling between them. 
In such cases, we shall say that $\alpha$ and $\beta$ are in the same
 {\it universality class}.
It is straightforward to see that if
\beq
          {N_{\alpha i}\over N_{\beta i}} ~=~ {N_{\alpha j}\over N_{\beta j}}
             ~~~~~~ {\rm for~all}~~(i,j)~, 
\label{preuniversalityclass}
\eeq
\ie, if the $\alpha$-row of the population matrix is a multiple of the $\beta$-row,
then $\alpha$ and $\beta$ will be in the same universality class.
Indeed, in such cases, the $\alpha$-characteristic need not be correlated with the 
probability deformations, but
the $\alpha$- and $\beta$-model spaces
nevertheless experience identical deformations.
Phrased slightly differently, this means that although the $\alpha$-model subspace
experiences a non-trivial deformation in passing to the corresponding probability space,
the $\beta$-model subspace experiences exactly the same deformation.

It turns out that Eq.~(\ref{preuniversalityclass}) is not the most general
condition which guarantees that $\alpha$ and $\beta$ are in the same universality class,
since we can also have 
situations in which there exist (subsets of) intrinsic probabilities $p_i$ such that
$p_i/p_k = p_j/p_\ell$.   
In such cases, we do not need to demand the strict condition in Eq.~(\ref{preuniversalityclass}),
but rather the more general condition
\beq
          {N_{\alpha i}\over N_{\beta k}} ~=~ {N_{\alpha j}\over N_{\beta \ell}}
             ~~~~~~ {\rm for~all~sets}~~(i,j,k,\ell)~~ {\rm for~which~~} 
        {p_i\over p_k} = {p_j\over p_\ell}~.
\label{universalityclass}
\eeq
We shall therefore take this to be our most general definition for when
two physical characteristics $\alpha$ and $\beta$ are in the same universality class.
However, it is easy to see that even when $p_i/p_k = p_j/ p_\ell$
has no solutions with $i\not=k$ and $j\not =\ell$, there will always exist
the trivial solution when $i=k$ and $j=\ell$.  In this case, Eq.~(\ref{universalityclass})
reduces back to Eq.~(\ref{preuniversalityclass}).

Regardless of the relations between the different physical characteristics $\alpha$,
the fundamental problem that concerns us can be summarized as follows.
As we construct model after model, we can keep a running tally of
$M_\alpha(D)$ for each relevant physical characteristic $\alpha$
(or for each relevant combined set of characteristics $\alpha$).
Equivalently, this information may be expressed as $M_\alpha(d_\alpha)$, where
we express the number of models $M_\alpha$ as a function of $d_\alpha$,
the numbers of draws which have yielded an $\alpha$-model regardless of whether
that model has not previously been seen.  
Ultimately, on the basis of this information,
our goal is to determine correlations between
these sets of characteristics across the entire landscape --- \ie,
we wish to determine the values of ratios such as $N_\alpha/N_\beta$,
where $N_\alpha\equiv \sum_i N_{\alpha i}$.
However, we now see that we face two fundamental hurdles:
\begin{itemize}
\item  First, it is not possible in practice to determine $N_\alpha/N_\beta$
       from the $\langle M_\alpha(D) \rangle$ or $\langle M_\alpha(d_\alpha)\rangle$ 
       because we do not have prior information about the 
        partial population matrix $N_{\alpha i}$
        or the individual probabilities $p_i$, both of which
        enter into Eq.~(\ref{bigeq}).  Indeed, even if we were
        willing to do a numerical fit and had sufficient statistical 
        data with which to conduct it,
       we do not even know the number 
         of distinct exponentials which enter into the sums in Eq.~(\ref{bigeq}),
         and it is always possible to improve accuracy (and thereby
          dramatically change the resulting best-fit values for the $N_{\alpha i}$)
         simply by introducing additional exponentials into the sum.
\item   Second, even if we could solve the mathematical problem
         of extracting $N_\alpha/N_\beta$ from $\langle M_\alpha(D)\rangle$
           or $\langle M_\alpha(d_\alpha)\rangle$,
           we do not know to what extent 
         $\langle M_\alpha(D)\rangle$ or 
         $\langle M_\alpha(d_\alpha)\rangle$ 
         can be taken to approximate
        the exact, discrete integers $M_\alpha(D)$ or $M_\alpha(d_\alpha)$ that are actually measured. 
       Clearly, we expect that this approximation should become
       very good as we explore sufficiently large portions of the corresponding
          entire model spaces, but 
         we cannot {\it a priori}\/ determine
        when this approximation might actually be valid
           because we do not know the absolute populations
         $N_{\alpha i}$ of these spaces. 
\end{itemize}

\begin{figure}
\centerline{
     \epsfxsize 4.0 truein \epsfbox {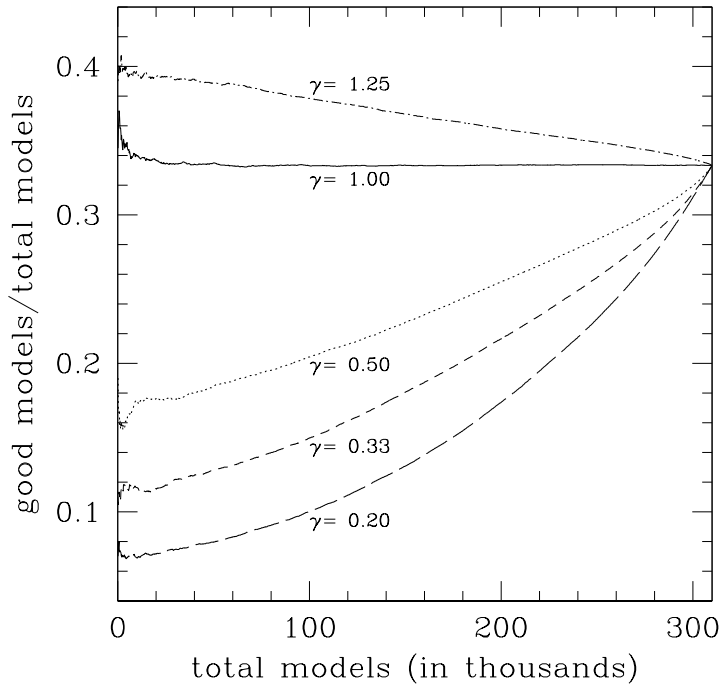}
    }
\caption{Results of an actual numerical simulation showing 
   how a correlation can ``float'' as the model space is explored.  
   For this simple example, we have assumed 
   only two disjoint populations
   of models with different intrinsic probabilities and likewise 
   assumed only two physical characteristics (denoted ``good'' and ``other'').  
   Moreover, we assumed a complete correlation
   between probabilities and physical characteristics, so that our
   $2\protect\times 2$ population matrix is diagonal.
   We assumed a model space consisting of 310,000
   models, one third of which are designated ``good'';  likewise,
   this simulation was run repeatedly with different probability ratios
   $\gamma\equiv p_{\rm good}/p_{\rm other}$ reflecting
   the intrinsic bias of our model-generation procedure.  
   Despite all of these simplifying assumptions,
   we see that our correlations float very strongly as a function
   of the sample size of distinct models found, 
   with the ratio of the numbers of ``good'' models
   to total models approaching the true value (=1/3) only when the total
   model space is explored.  We also see that our statistics from
   this random simulation do not follow any semblance of smooth behavior
   until we have examined at least 20,000 distinct models, representing
   approximately 6\% of the total model space.  }
\label{newSim} 
\end{figure}

Thus, it would clearly be an error to assume that $N_\alpha/N_\beta$ can
be identified as $M_\alpha(D)/M_\beta(D)$ for any particular value of $D$
(unless, of course, we have already saturated the model space, with $D\gg N_{\alpha,\beta}$).
Indeed, if we were to make this error, we would find that our proposed
ratio $N_\alpha/N_\beta$ would ``float'' --- \ie, it would evolve as a function of $D$.
This is, ultimately, the problem of floating correlations.
This behavior is illustrated in Fig.~\ref{newSim}, which shows the 
results of an actual simulation 
in the simple case in which there are only
two populations in each variable ($\alpha$ and $i$) and where the 
population matrix $N_{\alpha i}$ is diagonal.  Even in this dramatically simplified case,
we see that our observed ratios of models float dramatically as a function
of sample size,
reaching the true value only when the full model space has been reached.

Having described the problem in mathematical terms, we shall now
propose a solution.
The solution is relatively simple in principle, but its proper
implementation is somewhat subtle.  We shall therefore defer a discussion
of its implementation to the next section.

We shall begin by concentrating on the simplest case in which
the population matrix $N_{\alpha i}$ is diagonal.
In this case, all physical properties of interest
are perfectly correlated with the different probability deformations.
Thus, we can identify the $\alpha$-population with some value $i$, the
$\beta$-population with some value $j$,
and so forth.

Our goal is to generate a value representing $N_{\alpha}/N_{\beta}$ 
for some pre-determined (sets of) physical characteristics $\alpha$ and $\beta$.
However, all we can do is make repeated draws from the urn, slowly developing
tallies $M_{\alpha,\beta}(D)$ or $M_{\alpha,\beta}(d_{\alpha,\beta})$ 
of the distinct models in 
these respective classes.
As we continue in this process of drawing from the urn,
it becomes increasingly difficult to find new, hitherto-unseen models 
in each class.  Indeed, viewing these classes as entirely separate,
we see that the probability that each model 
selected from class $\alpha$ or class $\beta$ will not have previously been seen is
\beq
         P^{(\alpha)}_{\rm new}(d_\alpha)~=~ 1 - {M_\alpha(d_\alpha)\over N_\alpha}~,~~~~~
         P^{(\beta)}_{\rm new}(d_\beta)~=~ 1 - {M_\beta(d_\beta)\over N_\beta}~.
\label{probbs}
\eeq
These probabilities can be taken as measures of how fully a given
model space is explored.
Therefore, rather than attempt to identify 
\beq
         {N_\alpha\over N_\beta}  ~~{\stackrel{?}{=}}~~ {M_\alpha(D)\over M_\beta(D)}~
\eeq 
for any single value of $D$,
our solution is to instead identify
\beq
        \framebox{$\displaystyle
        ~{N_\alpha\over N_\beta}  ~=~ {M_\alpha(d_\alpha')\over M_\beta(d_\beta'')} 
             \Bigg |_{ P^{(\alpha)}_{\rm new}(d_\alpha') = 
                    P^{(\beta)}_{\rm new}(d_\beta'')  } 
            $~}
\label{kgolden}
\eeq
for two {\it different}\/ draw values $d_\alpha'$ and $d_\beta''$ 
which are chosen such that their respective production probabilities are equated.
Note that while the quantity $d_\alpha'$ will correspond to a certain total draw count $D'$,
the quantity $d_\beta''$ will generally correspond to a {\it different}\/ total draw count $D''$.
In other words, we do not extract the desired ratio $N_\alpha/N_\beta$ by comparing
$M_\alpha(D)$ and $M_\beta(D)$ at the same simultaneous point in the search process;
rather, we compare the value of $M_\alpha$ measured at one point in the search process
(\ie, after $D'$ total draws) with the value of $M_\beta$ measured at a {\it different}\/ point 
in the process (\ie, after $D''$ total draws).
As indicated in the condition in Eq.~(\ref{kgolden}), 
these different points are related by the fact
that {\it they correspond to points at which the corresponding $\alpha$- and $\beta$-model spaces
are equally explored}\/.
This then completely overcomes the biases that result from the fact that the different model
spaces are generally being explored at different rates. 

Of course, in the process of randomly generating string models, 
we cannot normally control whether a random new model is of the $\alpha$- or $\beta$-type.
Both will tend to be generated together, as part of the same random search.
Thus, if $D''>D'$, our procedure requires that we completely {\it disregard}\/
the additional $\alpha$-models that might have been generated in the 
process of generating the required, additional $\beta$-models.     
This is the critical implication of Eq.~(\ref{kgolden}).  
Rather than let our model-generating procedure continue for a certain
duration, with statistics gathered at the finish line, we must instead
establish two separate finish lines for our search process.  Of course, these finish lines
are arbitrary and must be chosen such their respective $\alpha$- and $\beta$-production 
probabilities  are equated.  However, these finish lines will not generally 
coincide with each other, which requires that some data actually be disregarded
in order to extract meaningful statistical correlations.

Thus far, we have been describing the situation in which we seek 
to obtain statistics comparing only two different groups of physical
characteristics $\alpha$ and $\beta$.  In general, however, we might wish to
compare whole sets of physical characteristics $\lbrace \alpha,\beta,\gamma,...\rbrace$.
Our procedure then requires that we establish a whole host of correlated finish lines,
one for each set of physical characteristics, and   
use Eq.~(\ref{kgolden}) to make pairwise comparisons.

Given Eq.~(\ref{probbs}), the result in Eq.~(\ref{kgolden}) follows quite trivially from the
condition that $P^{(\alpha)}_{\rm new}(d_\alpha')  = P^{(\beta)}_{\rm new}(d_\beta'') $.
The simplicity of this statement
may even seem to be a tautology, and indeed the difficulty in extracting the desired
correlation ratio $N_\alpha/N_\beta$ is now reduced to the practical question of determining
when the probability condition in Eq.~(\ref{kgolden}) is satisfied.  
This will be the focus of the next section. 
However, 
the important point is that we can overcome all of the biases inherent in the model-generation
process by focusing on the {\it probabilities}\/ for generating new distinct models,
and by comparing the numbers of models which have emerged at {\it different}\/ points in 
the model-generation process ---
 points at which these respective production probabilities are equal.

As indicated above, Eq.~(\ref{kgolden}) has been derived for the simple case in which 
the population matrix $N_{\alpha i}$ is diagonal.
However, {\it as long as $\alpha$ and $\beta$ are in the same universality class}\/,
it turns out that this result also holds for the more general case
in which our populations $\alpha$ and $\beta$ 
are non-trivially distributed across different probabilities $p_i$.
This statement is proven in the Appendix, and as we shall see below, this case
actually covers a large fraction of physically interesting characteristics.
Thus,  even in this case, we can overcome the biases inherent in the model-generation
process by focusing on the {\it probabilities}\/ for generating new distinct models
at different points in the model-generation process.

\section{Equating probabilities, and the uses of attempts/model}
\label{attemptspermodel}
\setcounter{footnote}{0}

The fundamental task that remains is to develop a method of
measuring the
restricted probabilities $P^{(\alpha,\beta)}_{\rm new}(d_{\alpha,\beta})$
which appear in Eq.~(\ref{kgolden}), or at least to develop 
a method of determining when these probabilities are equal.
At first glance, it might seem that this should be a relatively
simple undertaking.
Since we naturally generate data such as 
$M_{\alpha,\beta}(d_{\alpha,\beta})$ in the course of our model search,
it might seem that we could determine the individual
model-production probabilities simply by taking a derivative:
\beq
         P^{(\alpha)}_{\rm new}(d_\alpha) ~~{\stackrel{?}{=}}~~
              {{\rm d}\/ M_\alpha(d_\alpha)\over {\rm d}\, d_\alpha}~.
\eeq

Unfortunately, it turns out that taking such a derivative is 
computationally unfeasible.
The reason is that whereas the theoretical
expectation value 
$\langle M_\alpha(D)\rangle$ is a smooth, 
continuous function,
the actual ``measured'' quantities $M_\alpha(D)$ are necessarily 
discrete, jumping from integer to integer at unpredictable values of $D$
or $d_\alpha$.
Of course, one could perhaps extract $\langle M(D)\rangle$ by repeating
the same model-generation process over and over and averaging the results, 
but this is computationally expensive and redundant --- hardly an efficient
solution for a problem which has only arisen 
in the first place
because our computational power is already stretched to the maximum extent.

Indeed, the overall problem is that the production probabilities 
$P^{(\alpha,\beta)}_{\rm new}(D)$ --- which are the only  
true legitimate measure of the degree to which a model
space is explored --- fail to be a computationally {\it practical}\/ 
measure 
because they are extremely sensitive to the 
difference between $M_\alpha(D)$ and $\langle M_\alpha(D)\rangle$. 
What we require, by contrast, is an alternative measure of the extent
to which a given model space is explored,
a measure which may be only approximate but which is less sensitive
to the difference between $M(D)$ and $\langle M(D)\rangle$
and which can therefore be implemented in an actual automated
search through the model space.
 
To get an idea how to proceed, let us begin by considering 
the simplified case in which the population matrix $N_{\alpha i}$ is
actually diagonal.  In this case, all physical properties of interest
are perfectly correlated with the different probability deformations,
so that we can identify the $\alpha$-population with some value $i$, the
$\beta$-population with some value $j$,
and so forth.
Thus we expect $\langle M_\alpha(D)\rangle$ (or equivalently
$\langle M_i(D)\rangle$ for some $i$)
to follow Eqs.~(\ref{otherprobinewmodel})
and (\ref{probinewmodel2}).
Given these equations, it then follows that 
\beq 
        {N_i\over N_j} ~=~ {M_i(D')\over M_j(D'')}
\label{want}
\eeq
only when we satisfy the balancing condition
\beq
         { D' p_i \over \Omega_{\rm prob} } ~=~ 
         { D'' p_j \over \Omega_{\rm prob} }  ~
\label{balancing1}
\eeq
or equivalently
\beq
         { \langle d'_i\rangle  \over N_i } ~=~ { \langle d''_j\rangle  \over N_j } ~.
\label{balancing1new}
\eeq

Since we do not know the values of the $p_i$ or the $N_i$,
it is not possible to determine the balanced pairs of values $(D',D'')$
or $(\langle d'_i\rangle, \langle d'_j\rangle)$ using these equations.
However, since Eq.~(\ref{balancing1}) implies Eq.~(\ref{want}), we
can multiply each side of Eq.~(\ref{balancing1new}) by $N_i /\langle M_i(d'_i)\rangle $ 
or $N_j /\langle M_j(d''_j)\rangle $ respectively 
to obtain the equivalent balancing condition
\beq
         { \langle d'_i\rangle  \over \langle M_i(d'_i) \rangle } ~=~
         { \langle d''_j\rangle  \over \langle M_j(d''_j) \rangle } ~.
\label{balancing2}
\eeq
Unlike Eq.~(\ref{balancing1}), this balancing equation is easy to interpret 
and implement in a computer search since $\langle d_i / M_i(d_i)\rangle$ is
nothing but the expectation value of the ratio of `attempts' to `models',
where `attempts' refers
to the total number of $i$-models drawn and `models' refers to the total
number of actual {\it distinct}\/ $i$-models drawn.  
Thus, we can view our balancing condition as one which equates {\it cumulative attempts per model}\/,
where our attempts are restricted to those which yielded a model (whether distinct or not)
in the appropriate class.

\begin{figure}[htb]

\centerline{
   \epsfxsize 3.7 truein \epsfbox{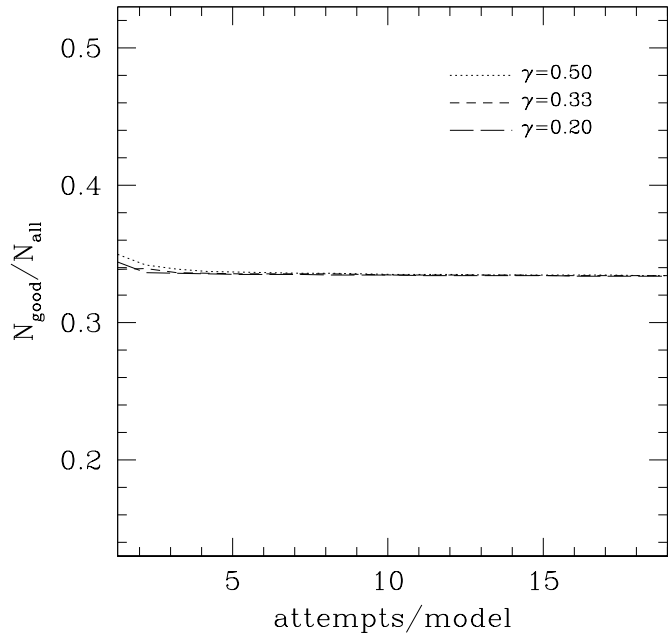}
 }
\caption{Results of a numerical simulation involving the same setup as
in Fig.~\ref{newSim}, but now with $N_{\rm good}/N_{\rm all}$ extracted
through Eq.~(\ref{kgolden2}) and plotted as a function of the value of
            $d'_{\rm good}/M_{\rm good}(d'_{\rm good})=
             d''_{\rm other}/M_{\rm other}(d''_{\rm other})$.
As we see, use of this method enables us to extract the correct value
$N_{\rm good}/N_{\rm all} = 1/3$ 
with considerable accuracy even for relatively small values of attempts/model,
regardless of the value of the bias $\gamma$.  We also ran similar simulations
in which each of the models in the model space was subjected to an additional arbitrary
probability deformation;  as long as the condition in Eq.~(\ref{universalityclass})
was enforced, the resulting plot remained essentially unchanged.  }
\label{NewMethod} 
\end{figure}

Of course, it may initially seem that attempts/model is no better than production probabilities
since they are both essentially equivalent when the population matrix is diagonal or has rescaled rows.
However, the important point is that since attempts/model does not involve a derivative of $M(D)$, 
this quantity is actually less sensitive to
the difference between $M(D)$ and $\langle M(D)\rangle$ than the production 
probabilities $P_{\rm new}^{(\alpha)}$.
Thus, we may replace
\beq
         { \langle d'_i\rangle  \over \langle M_i(d'_i) \rangle } ~\longrightarrow~
         {d'_i \over M_i(d'_i)}~
\label{replace}
\eeq
in Eq.~(\ref{balancing2}) 
without seriously damaging our ability to extract the desired ratio $N_i/N_j$ 
(or $N_\alpha/N_\beta$).
This fact is illustrated in Fig.~\ref{NewMethod}.

This result is valid for the case when the population matrix is diagonal.
However, it is straightforward to see that these results also hold 
for any $\lbrace \alpha,\beta, ...\rbrace $ which are in the same universality class
[as defined in Eq.~(\ref{universalityclass})].
Because the probability spaces corresponding to models with each of these
characteristics have identical deformation patterns, 
we can repeat the above derivation and find that attempts/model continues
to be a fairly accurate measure parametrizing the degree to which a given model space
is explored.
As it turns out, many physical characteristics of interest 
$\lbrace \alpha,\beta,...\rbrace$ have the property that they share identical probability deformations
for a given model-construction formalism,
and are thus in the same universality class. 
Thus, for these characteristics, attempts/model can be used in place of production probabilities
in extracting population ratios:
\beq
        \framebox{$\displaystyle
        ~{N_\alpha\over N_\beta}  ~=~ {M_\alpha(d'_\alpha)\over M_\beta(d''_\beta)} 
             \Bigg |_{  
                  {d'_\alpha\over M_\alpha(d'_\alpha)} =            
                  {d''_\beta\over M_\beta(d''_\beta)} } 
            $~}~.
\label{kgolden2}
\eeq
Indeed, we can ``experimentally'' verify whether our chosen physical characteristics
$\alpha$ and $\beta$ are in the same universality class by calculating the ratio $N_\alpha/N_\beta$
as a function of the chosen number of attempts/model using this relation, and verifying that
this ratio does not experience any float as a function of attempts/model.
The absence of any float indicates that the physical
characteristics $(\alpha,\beta)$ are in the same universality class.
We shall see explicit examples of this   
situation in Sect.~\ref{warning}.

One important cross-check is to verify that
\beq
        {N_{\alpha}\over N_{\beta}} ~=~
        {N_{\alpha}\over N_{\gamma}} \cdot
        {N_{\gamma}\over N_{\beta}}
\label{crosscheck}
\eeq
for all $(\alpha,\beta,\gamma)$ in the same universality class, 
where each of these fractions is individually extracted through Eq.~(\ref{kgolden2}).
Since Eq.~(\ref{crosscheck}) is not guaranteed to hold on the basis of the
definition in Eq.~(\ref{kgolden2}), its validity provides an important check
on any results we obtain.

It is important to note that this procedure only yields a set of {\it relative}\/
abundances of the form $N_{\alpha}/N_{\beta}$ within the same universality class.
This is usually the best one can do.
However, it is occasionally possible to convert this information
to absolute proportions of the form $N_\alpha/ N_{\rm all}$.
For example, if the characteristics $\lbrace \alpha,\beta,\gamma,...\rbrace$ 
are non-overlapping, all in the same
universality class, and happen to span the entire
space of possible physical characteristics,
then $N_{\rm all}=N_\alpha+N_\beta+N_\gamma +...$ and we can therefore extract
the absolute probabilities:
\beq
           \Omega_{\alpha} ~\equiv~ {N_\alpha\over N_{\rm all}}~=~
           {N_\alpha\over N_\alpha + N_\beta+ N_\gamma+ ...}
                 ~= ~ \left( 1 + {N_\beta \over N_\alpha} + {N_\gamma\over N_\alpha} + ...\right)^{-1}~.
\label{Omegaalpha}
\eeq
Alternatively, we can sometimes avoid this procedure by simply letting
$\overline{\alpha}$ denote the complement of $\alpha$
(\ie, the characteristic that a given model does {\it not}\/ contain the characteristic
associated with $\alpha$) and calculate $N_\alpha/N_{\overline{\alpha}}$.
If $\alpha$ and $\overline{\alpha}$ happen to be in the same universality class,
then our result for $N_\alpha/N_{\overline{\alpha}}$ will be stable
and $\Omega_\alpha$ can then be extracted through Eq.~(\ref{Omegaalpha})
where we identify $N_{\overline{\alpha}} = N_\beta+N_\gamma+ ...$. 
Calculating $\Omega_\alpha$ for one member $\alpha$ of a given  
universality class will then enable us to obtain $\Omega_\beta$ for
every other member $\beta$ of the class.
However, we stress that this relies on the assumption 
that $\alpha$ and $\overline{\alpha}$ are in the same universality class,
a situation which is not guaranteed to be the case.

Finally, of course, we may face the most general situation     
in which two physical characteristics $\alpha$ 
and $\beta$ are not in the same universality class.
In such cases, even the ratios $N_\alpha/N_\beta$ determined through 
Eq.~(\ref{kgolden2})
will float as a function of the number of attempts/model.
Indeed, as mentioned above, this provides a test (indeed, the only viable test) of whether  
two physical characteristics $\alpha$ and $\beta$ are truly in the same
universality class.
However, even if $\alpha$ and $\beta$ are not in the same universality
class, it may nevertheless be possible to extract individual absolute probabilities $\Omega_\alpha$
and $\Omega_\beta$ through Eq.~(\ref{Omegaalpha}) if $\alpha$ and $\beta$ are
in the same universality classes as $\overline{\alpha}$ and $\overline{\beta}$ respectively.
We can then indirectly calculate the relative probability
$N_\alpha/N_\beta =  \Omega_\alpha/\Omega_\beta$.

Even when $\lbrace \alpha,\beta,\gamma\rbrace$ are not in the same universality
class, the cross-check in Eq.~(\ref{crosscheck}) must continue to hold.
However, each individual fraction will not be stable as a function of
attempts/model unless it is determined indirectly
through $\Omega_{\alpha,\beta,\gamma}$.
For example, let us imagine that $\alpha$ and $\beta$ are in the
same universality class but $\gamma$ is in a different universality class.
In this case, we can use Eq.~(\ref{kgolden2}) to obtain each of the fractions
in Eq.~(\ref{crosscheck}), but the two factors on the right side of
Eq.~(\ref{kgolden2}) will float as a function of attempts/model, constrained
only by the requirement that their product be fixed.
However, determining these factors through their absolute $\Omega$-probabilities 
will enable stable results to be reached.

\begin{figure}
\centerline{
\epsfxsize 2.9 truein \epsfbox{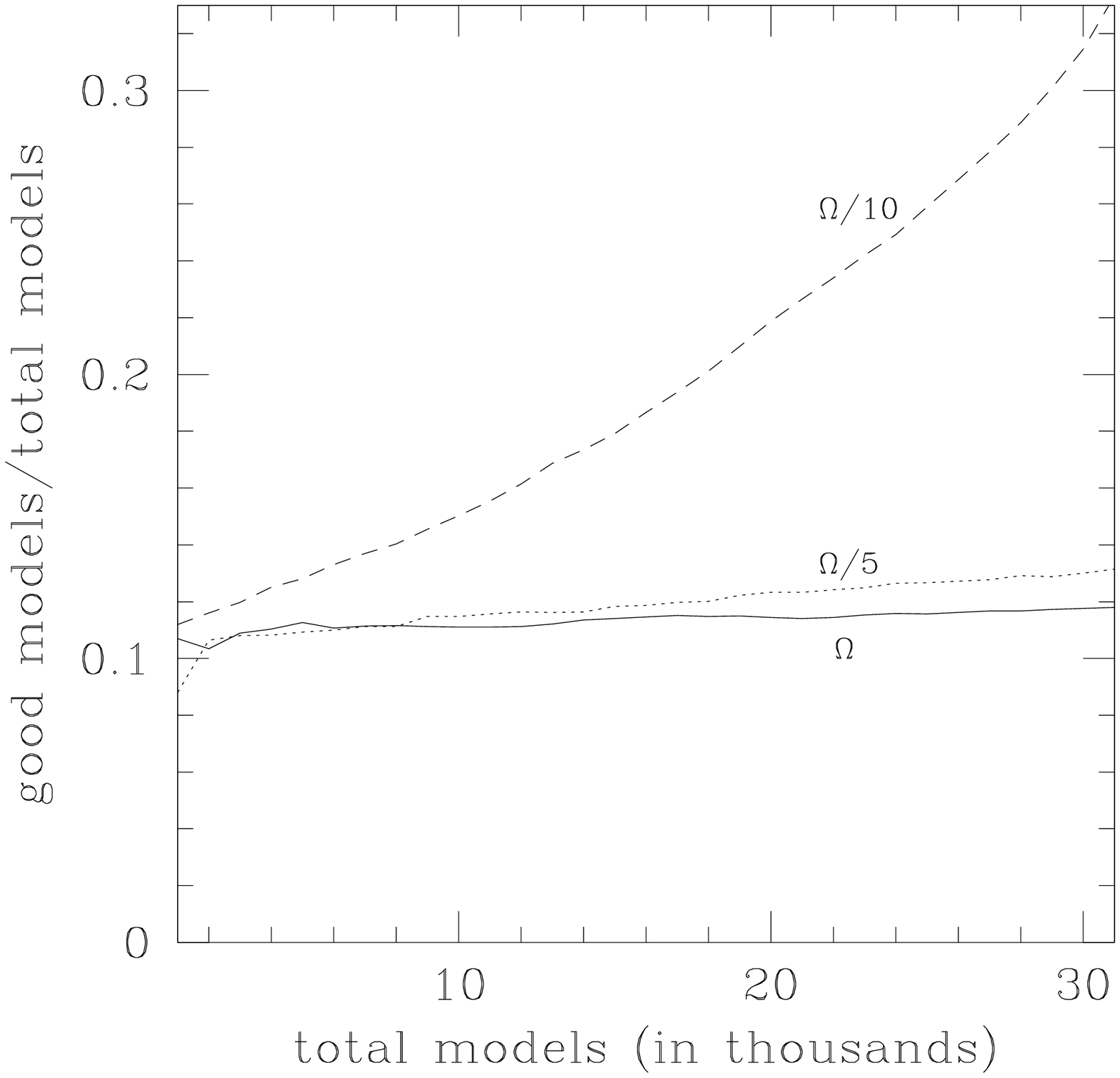}
\hfill
\epsfxsize 2.9 truein \epsfbox{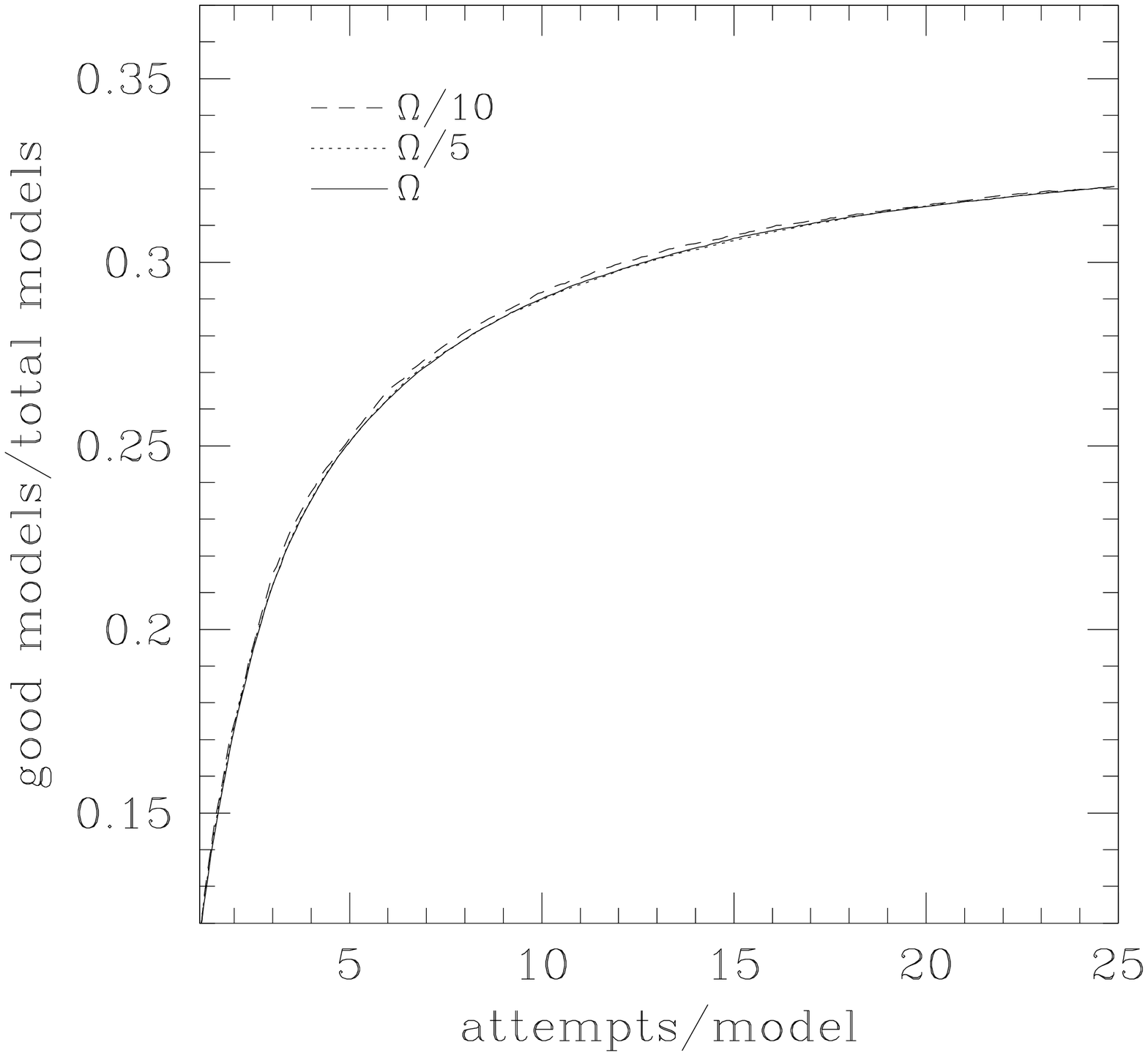}
 }
\centerline{
\epsfxsize 2.9 truein \epsfbox{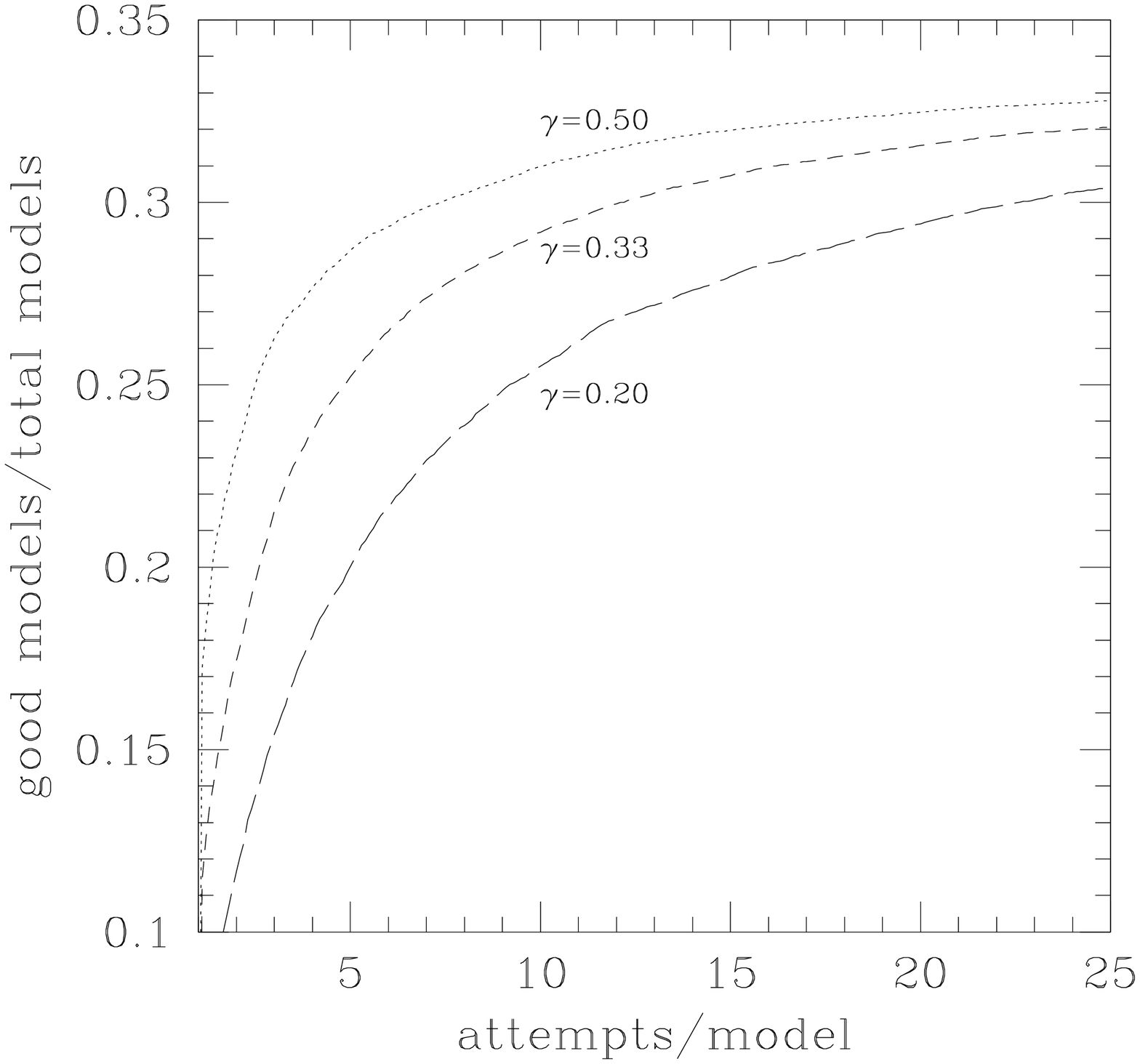}
 }
\caption{{\it Top figures:}\/  
Two different ways of presenting the same simulation results.  In each graph, 
we have plotted the results from three simulations
in which the number of models designated ``good'' (according to a pre-defined characteristic)
is one-third of the total.  However, these simulations differ in the total size
of the model space, with total model-space volumes taken to be $\Omega$, $\Omega/5$, and $\Omega/10$ 
where $\Omega=310000$.
Each simulation also incorporated a model-generation 
probability bias $\gamma\equiv p_{\rm good}/p_{\rm other}=1/3$.
We see that the observed ratio $N_{\rm good}/N_{\rm all}$ floats in each case, but 
plotting these results as a function of sample size (as in the left figure)
does not allow us to separate the effects of
the bias from the effects of the different volumes.  However, 
plotting these results in terms of 
attempts/model (as in the right figure) enables us to 
completely eliminate the effects of the differing model-space volumes.
 {\it Bottom figure:}\/  The results of the same simulation,
but now with differing bias ratios $\gamma$.  We see that
unlike the effects of differing volumes,
the effects of bias cannot be overcome simply by considering model spaces
at similar levels of exploration.}
\label{bias} 
\label{sampsize} 
\end{figure}

We see, then, that our solution to the problem of floating correlations
involves more than simply tallying the populations of different models
generated in a random search --- it also requires information about
 {\it how}\/ they were generated, and in particular how many {\it attempts}\/
at producing a distinctly new model are required before a given such model
is actually found.  While this represents new data which might 
not otherwise have received any special attention,
we see that it is this new ingredient which enables us to evaluate the
degree to which a given model space has been explored.
Moreover, it is relatively easy to keep track of this 
information during the model-generation process.

In this connection, it is important to note that 
attempts/model can also have additional important uses beyond Eq.~(\ref{kgolden2}).
For example, 
attempts/model can be used as
a measure of the extent to which a given model space has been explored ---
even in the presence of an unknown model-generation bias. 
Thus, use of attempts/model
can allow comparisons between model spaces of different (unknown) sizes.
This property is illustrated in Fig.~\ref{sampsize},
which shows that use of attempts/model can completely eliminate the effects 
of differing model-space volumes. 
However, it is clear from these plots 
that the existence of model-generation bias can continue to make 
a determination of $N_{\rm good}/N_{\rm other}$ impossible until the model 
space is nearly fully explored.  Indeed, we see that the 
amount of model space which must be explored in order
to overcome the bias depends on 
the value of $\gamma$.

\section{A heterotic example}
\label{warning}
\setcounter{footnote}{0}

In this section, we shall illustrate 
the above ideas and their implementation in an actual example drawn from the
heterotic string landscape.
As we shall see, the use of these ideas leads
to correlations that differ markedly from those which would have na\"\i vely 
been apparent from only a partial data set.

The models we shall examine are all four-dimensional perturbative heterotic 
string models with ${\cal N}=1$ spacetime supersymmetry, 
formulated through through the free-fermionic construction~\cite{KLT}.
In the language of this construction,
worldsheet conformal anomalies are cancelled through the introduction of free 
fermions on the worldsheet, and  
different models are realized by varying (or ``twisting'') the boundary conditions of
these fermions around the two non-contractible loops of the
worldsheet torus while simultaneously varying
the phases according to which the contributions of each such
spin-structure sector are summed in producing the one-loop
partition function.
For the purposes of our search, all worldsheet fermions were taken to be complex
with either Neveu-Schwarz (anti-periodic) or Ramond (periodic) boundary conditions.
However, we emphasize that
alternative but equivalent languages for constructing such models exist.
For example, we may bosonize these worldsheet fermions and
construct ``Narain'' models~\cite{Narain,Lerche} in which the resulting complex worldsheet
bosons are compactified on internal lattices of appropriate dimensionality
with appropriate self-duality properties.
Furthermore, many of these models have additional geometric realizations
as orbifold compactifications with randomly chosen
Wilson lines;  in general, the process of orbifolding
is quite complicated in these models, involving many sequential overlapping 
layers of projections and twists.

A full examination of these statistical correlations for such ${\cal N}=1$
string models will be presented in Ref.~\cite{next}.  Indeed, many of the
techniques behind our model-generation techniques and subsequent statistical analysis
are similar to those described in Ref.~\cite{dienes}.
However, our goal here is merely to provide an example of  
how certain statistical correlations float, and how stable results can 
nevertheless be extracted.

Towards this end, we shall restrict our attention to a simple question:
with what probabilities do certain gauge-group factors appear in the total 
(rank-22) gauge group of such ${\cal N}=1$ string models?
To address this question, we randomly constructed a set of 
$\approx 3.16$~million distinct models in this class.
This set of models is 25 times larger than that examined in Ref.~\cite{dienes},
and thus represents the largest set of distinct heterotic string models which have ever
been constructed to date.  
We emphasize that the distinctness of these models is measured, as discussed in Sect.~2, 
on the basis of their resulting physical characteristics in spacetime and
not on the basis of the internal worldsheet 
parameters from which they are derived.

One feature which is immediately apparent from such models is that while $U(1)$ and
$SU(2)$ gauge-group factors are fairly ubiquitous, $SU(3)$ gauge-group factors are 
relatively rare.  Indeed, if we restrict our attention to the first $1.25$~million
models that were generated in this set, 
we find that more than $90\%$ of these models 
exhibit at least one $U(1)$ or $SU(2)$ gauge-group factor, while less than $\approx 50\%$ 
of these 
models exhibit an $SU(3)$ gauge-group factor.
Thus, we have what appears at first glance to be a striking 
disparity:  $SU(3)$ gauge-group factors appear to be significantly less likely
to appear than $U(1)$ or $SU(2)$ gauge-group factors, at least in this 
perturbative heterotic corner of the landscape. 
 
However, an alternative explanation might simply be that our model-construction
technique (in this case, one involving free worldsheet complex fermions with
only periodic or anti-periodic worldsheet boundary conditions)
may have certain inherent tendencies to produce
models with $U(1)$ or $SU(2)$ gauge-group factors more easily than
to produce models with $SU(3)$ gauge-group factors.
Indeed, even though this construction technique may ultimately be capable
of producing {\it more}\/ models with $SU(3)$ gauge-group factors than
$U(1)$ or $SU(2)$ gauge-group factors (thereby causing the $SU(3)$ models to occupy a larger
relative volume of the associated {\it model}\/ space), it may simply be that
the models with $SU(3)$ gauge-group
factors may be more {\it difficult to reach}\/ and thus occupy a smaller
volume within the associated {\it probability}\/ space.
If this is true, then we cannot hope to reach any conclusion about the relative
abundances of $U(1)$, $SU(2)$, and $SU(3)$ gauge-group factors on the basis
of a straightforward census of the models we have generated.

Again, we emphasize that this is not a problem unique to the 
free-fermionic construction.  Literally any construction procedure
will have an intrinsic bias towards or against certain string models,
yet this need not have anything to do with the ultimate statistical
properties across the corresponding model spaces. 
Thus, since we can examine at best only a necessarily finite sample of models,  
it is clear that we are not able to extract
any meaningful information from a census study of a finite model sample alone.

\begin{figure}[htb]
\centerline{\epsfxsize 4.0 truein \epsfbox{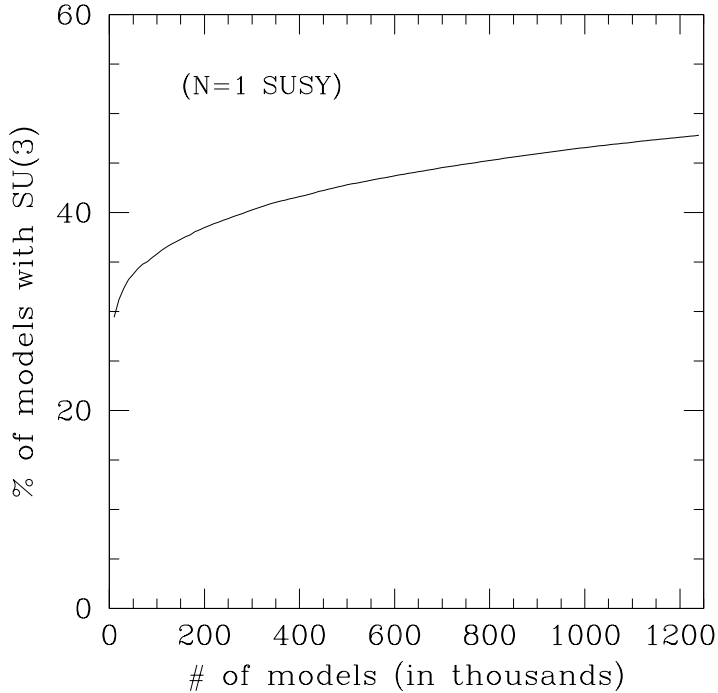}}
\caption{The percentage of distinct four-dimensional 
         ${\cal N}=1$ supersymmetric heterotic string models 
  exhibiting at least one $SU(3)$ gauge-group factor, plotted as 
  a function of the number of models examined for the first $1.25$~million 
   models.  We see that as we generate further models, 
   $SU(3)$ gauge-group factors become somewhat more ubiquitous --- 
    {\it i.e.}\/, the fraction of models with this property {\it floats}.    
   This implies that models with $SU(3)$ gauge-group factors occupy 
   a smaller relative volume within the probability space of models 
   than within the true model space itself.  One must therefore correct for 
    this distortion of the probability space relative to the model space 
   before any conclusions concerning
    the relative probabilities of specific gauge-group factors can be drawn.}
\label{EvolveSU3} 
\end{figure}

One clue that we are indeed dealing with a model-construction bias
in this example comes from examining the percentage of models exhibiting
an $SU(3)$ gauge-group factor {\it as a function of the number of distinct models we 
generated at different points in our search}\/.
This data is plotted in Fig.~\ref{EvolveSU3} 
for the first $1.25$~million models, 
and it is immediately clear that 
the percentage of models with $SU(3)$ gauge-group factors
 {\it floats}\/ rather significantly as a function of the sample size.
This implies that
  models with $SU(3)$ gauge-group factors occupy a smaller relative volume within
  the probability space of models than within the true model space itself.
We emphasize that this need not have been the case:
it could have turned out that gauge groups were uniformly 
distributed among the populations of models with different probabilities of production.  
However, 
by examining gauge-group correlations as a  function of the number of models generated,
we now have clear evidence that this is not the case.
Therefore,
before we can draw any conclusions concerning
the relative probabilities of specific gauge-group factors for such string models,
we must compensate for this 
distortion of the probability space relative to the model space.

At first glance, it might seem from Fig.~\ref{EvolveSU3} that 
the proportion of models with $SU(3)$ gauge-group factors appears to be saturating
somewhere near $50\%$ or $60\%$.
However, we must remember that the true size of the 
full model space is unknown.  This means that even 
though the proportion of models with $SU(3)$ seems to be floating very slowly,
it is difficult to judge how long this floating might continue if we were able to 
examine more models.
Even a small degree of floating could accumulate into a large change in the apparent
frequency of $SU(3)$ gauge factors.
Moreover, we are generally concerned with {\it correlations}\/ --- \ie, relationships
between two or more different variables.
For example, we might be concerned with a correlation between the appearance of an
$SU(3)$ gauge-group
factor and the appearance of an $SU(2)$ gauge-group factor or a $U(1)$ gauge group factor.
Since each individual gauge-group factor may experience its own degree of floating,
the net float of the correlation can be quite strong even if these individual
floats are rather weak.

In order to address these difficulties, we can therefore employ the methods outlined
in the previous section.  For example, we can let $\alpha$ represent the physical
characteristic that a string model contains a $U(1)$ gauge-group factor, $\beta$
represent the same for $SU(2)$, $\gamma$ the same for $SU(3)$, and so forth. 
If these $\alpha$, $\beta$, and $\gamma$ characteristics are in the same universality
class [as defined in Eq.~(\ref{universalityclass})], we can use
Eq.~(\ref{kgolden2}) directly to extract $N_\alpha/N_\beta$, $N_\beta/N_\gamma$, and so forth.  
Indeed, calculating these ratios as a function of attempts/model, we can verify whether 
$\alpha$, $\beta$, are $\gamma$ are truly in the same universality class. 
Moreover, even when these characteristics are not in the same universality class,
we can use the method outlined in Eq.~(\ref{Omegaalpha}) to obtain absolute probabilities
$\Omega_\alpha$ when $\alpha$ and $\overline{\alpha}$ are in the same universality class.  
In such cases, we can then convert all of our final information to 
the same absolute scale $\Omega_\alpha$.

Our results are shown in Table~\ref{example}.
For each listed gauge-group factor, we list 
the percentage of models containing this factor at least once 
(tallied across our sample consisting of the first $1.25$~million distinct
four-dimensional ${\cal N}=1$ heterotic string models we generated)
as well as the 
percentage to which this sample result ultimately ``floats'',
as extracted through Eqs.~(\ref{kgolden2}) and (\ref{Omegaalpha}).
Although not directly evident from the entries in this table,
it turns out from our analysis that each of these group factors
is in the same free-fermionic universality class, at least
as far as we can determine numerically.
Moreover, we were able to verify (again within numerical error)
that $\alpha$ and $\overline{\alpha}$ are in the 
same universality class for the case when $\alpha$
represents the $SU(5)$ characteristic.
It was through this observation that we were able to convert
the relative probabilities $N_\alpha/N_\beta$ into 
the absolute probabilities $\Omega_{\alpha,\beta}$
quoted in Table~\ref{example}.

\begin{table}[htb]
\begin{center}
\begin{tabular}{||c||c|c||}
\hline
group& finite sample & extracted $\Omega_{\alpha}$  \\
\hline
\hline
$U_1$ & 99.94& 95.6 \\
\hline
$SU_2$& 97.44 & 98.2 \\
\hline
$SU_3$ & 47.84 & 97.6 \\
\hline
$SU_4$&  51.04 &   29.5 \\
\hline
$SU_5$&  7.36 &  41.6\\
\hline
$SU_{>5}$ & 6.60 & 1.72 \\
\hline
$SO_8$ & 13.75 &  1.53  \\
\hline
$SO_{10}$ & 4.83 &  0.21 \\
\hline
$SO_{>10}$ & 2.69 &  0.054 \\
\hline
$E_{6,7,8} $ & 0.27 &  0.023 \\
\hline
\end{tabular}
\end{center}
\caption{Percentage of four-dimensional ${\cal N}=1$ supersymmetric heterotic 
  string models containing various gauge-group factors at least once in their
  total gauge groups.    
  Here $SU_{>5}$ indicates the appearance of {\it any}\/ $SU(n>5)$ factor, 
  while $SO_{>10}$ indicates any 
  $SO(2n)$ group with $n\geq 6$ and $E_{6,7,8}$ signifies any of the `E' groups.
  For each gauge-group factor, the `sample' column indicates to the percentages 
  of models exhibiting this factor
  across our sample of more than one million distinct 
  models in this class.  By contrast, the $\Omega_\alpha$ column lists the 
  corresponding values to which these percentages would ``float'', 
  as extracted through Eqs.~(\ref{kgolden2}) and/or (\ref{Omegaalpha}).  It is 
  clear that correcting
  for such probability deformations can result in abundances
  which are markedly different from those which appear within 
  a finite sample.}
\label{example}
\end{table}

As is evident from Table~\ref{example}, the effects of such floating
can be rather significant, resulting in relative percentages $\Omega_\alpha$ 
which often differ significantly from 
the percentages which are evident in only the finite sample set.
Perhaps the most significant example of this can be found 
in the relation between the $SU_5$ and $SU_4$ columns in Table~\ref{example}.  
At relatively low levels of exploration, one would easily conclude
that $SO(2n)$ groups (such as $SU_4 \sim SO_6$) are more common than $SU(n \ge 3)$
groups, since every $SO(2n)$ group has a higher probability of occurring than the
corresponding $SU(n)$ group of the same rank.  However, when the full model space is
extracted, it is clear that actually the reverse is true:
the `SU' groups actually dominate the model space even though they do not
dominate the probability space.
Indeed, the apparent paucity of `SU' groups in our finite sample
indicates nothing more than their difficulty of construction --- a feature
which is completely unrelated to their
overall abundance within this class of string models.
We see, then, that the issue of floating correlations can be 
rather important in any attempt to obtain statistical correlations 
through examination of only a finite data set.

We emphasize that although the procedures outlined in the previous
section are fairly robust, there can be numerous numerical/computational
difficulties which can cloud or obscure these results.  For example,
we found that it was much more difficult to extract information concerning
the $SU(3)$ gauge-group factor than for almost any other factor. 
We attribute this to the fact that the $SU(3)$ gauge-group characteristic
is predominantly distributed amongst models with extremely small intrinsic
probabilities $p_i$ in this construction, making it difficult
to reach significant penetration into this set  
with sufficiently large values of attempts/model.
Moreover, as we have stressed in Eq.~(\ref{replace}),
the actual numbers of attempts/model, just like the actual
numbers of models generated, are only approximations
to their mathematical expectation values. 
When attempting to extract correlations between models whose $p_i$
are of hierarchically different sizes, these numerical issues can
become severe. 
These numerical issues must therefore be dealt with on a case-by-case
basis when attempting to extract correlations from the landscape.

Given the results in Table~\ref{example}, one might wonder why we did not quote joint probabilities for
the composite Standard-Model gauge group $G_{\rm SM}\equiv SU(3)\times SU(2)\times U(1)$
or the composite Pati-Salam gauge group $G_{\rm PS}=SO(6)\times SO(4)$ in Table~\ref{example}.
The reason is that these composite groups $G_{\rm SM}$ and $G_{\rm PS}$ do not appear
to be in the same universality classes as their individual factors.
This, coupled with the numerical difficulties of dealing with apparently
small $p_i$, makes an analysis for these cases significantly more intricate.
The results for these cases will be given in Ref.~\cite{next}.

\section{Discussion}
\setcounter{footnote}{0}

In this paper, we have investigated some of the issues which challenge attempts to 
randomly explore the string landscape.  We identified an important generic
difficulty --- the problem of ``floating correlations'' --- and presented a method for overcoming
this difficulty which is applicable in a large variety of cases.
Moreover, we found that properly compensating for these floating correlations
can lead to statistical results which differ, in many cases
substantially, from the results which would have emerged 
from direct statistical examination of only a partial data set. 
We therefore believe that recognition of and compensation for these effects
are absolutely critical, and must play a role in any future string landscape study 
which operates through a random generation of string models.

It is worth emphasizing that this entire difficulty ultimately 
stems from our underlying ignorance of the properties of the
functions (discussed in Sect.~2) which map internal string-construction parameters  
into spacetime physical observables.  
If we had an explicit and usable representation for these functions,
we could avoid this whole problem completely since we could analytically (or computationally)
account for this kind of bias directly in our model-generating process.
It is only because of the difficulty of analyzing such functions in a general
way that we are forced into situations in which our model spaces 
experience such significant probability deformations.
These sorts of concerns also fail to play a role in various {\it field-theoretic}\/
analyses of the landscape~\cite{fieldtheory}.

It is also worth emphasizing that
although we have focused in this paper on the specific problem 
of surveying string models in a way suitable for string
landscape studies,
the mathematical problem we have been dealing with is actually far more general,
arising in all generic situations in which we seek to scan one space (such as the model space) while 
we only have direct computational access to a second space (such as the probability space)
whose relations to the first space are generally unknown or difficult to analyze analytically.
Thus, we expect our approach to this problem to have general applicability as well.  

Despite these facts, there are still many issues which are left unresolved by 
our methods.
Some of these issues are numerical and computational --- 
for example, one must develop techniques of overcoming other sorts of numerical 
instabilities and fluctuations which transcend the bias issues we have been discussing,
but which nevertheless can be significant.
Other issues are more abstract and mathematical ---
for example,
one must eventually develop new and efficient methods of 
generating and classifying string vacua.
One also requires additional theoretical input 
into the all-important question of determining which measure
is ultimately 
the most appropriate for landscape calculations. 
Finally, other issues are more detailed and potentially intractible ---
for example, although we have given a procedure for extracting statistical
correlations between physical observables in the same universality class,
we have not provided any procedure for relating physical observables in
different universality classes.
Barring successful resolution, all of these are critical issues which 
will likely hamper future statistical studies of the string landscape.

There are also other challenges which are inherent to all attempts at 
statistical explorations of the string landscape, be they numerical or analytic,
randomized or systematic.
Although discussed elsewhere
(see, \eg, Ref.~\cite{dienes}), we feel that they bear repeating because of 
their generality.

One of these has been termed the  ``G\"odel effect'' ---
the danger that no matter how many conditions (or input ``priors'')
one demands for a phenomenologically realistic string model,
there will always be another observable
for which the set of realistic models will make differing predictions.
Therefore, such an observable will remain beyond our statistical
ability to predict.
(This is reminiscent of the ``G\"odel incompleteness theorem''
which states that in any axiomatic system, there is always
another statement which, although true, cannot be deduced purely from the axioms.)
Given that the full string landscape is very large, consisting of perhaps
$10^{500}$ distinct models or more,
the G\"odel effect may represent a very real threat 
to our ability to ultimately extract true phenomenological predictions from
the landscape.

Another can be called the ``bulls-eye'' problem --- 
the realization that 
since we cannot be certain how our low-energy world is ultimately
embedded into a string framework,
we do not always know physical characteristics our ``target'' 
string models should possess.
For example, we do not know
whether our world becomes supersymmetric as we move upwards in
energy, or whether strong-coupling effects develop which
completely change our perspective on microscopic physics.  We do
not know whether our world remains essentially four-dimensional as we
move upwards towards the string scale, or whether there 
exist extra spacetime dimensions (large or small, flat or warped) 
which become evident at intermediate scales.
Indeed,
it is possible that nature might pass through many layers
of effective field theories at higher and higher energy scales before reaching an
ultimate string-theory embedding.  Absence of knowledge concerning the appropriate
string-theory embedding 
thereby limits our ability to identify {\it which}\/ 
statistical information about the
string landscape is the most important to extract.

A third challenge can be termed the ``lamppost'' effect ---
the danger of restricting one's attention to only those portions
of the landscape where one has control over calculational
techniques.
Ultimately, barring a complete classification of all consistent
string vacua, there is always the danger that there exists
a huge sea of unexplored string models whose properties are sufficiently
novel that they would invalidate 
any statistical conclusion we might have already reached.
This danger exists regardless of how detailed or
comprehensive an analysis we may have just performed.
Indeed, at any moment in time, our knowledge of string theory and various 
constructions leading to consistent string models is, by necessity,
quite limited.  A decade ago, one would have considered the heterotic
strings alone to have comprised the set of phenomenologically viable
string models.  The advent of the second superstring revolution has
opened the doorway to studies of Type~I strings, and recent realizations
concerning flux vacua have led to new ideas concerning moduli stabilization.
It is impossible to predict what the future might hold, and thus
it might be argued that any statistical analysis of known vacua is
at best premature.

Closely related to this is the problem of unknowns --- even within
a given string construction.
For example, the methods we have been describing in this paper
for sampling string models randomly can eventually allow us to evaluate,
with some certainty, how large a volume of the probability space 
of models might still have been missed in our search. 
However, although such a statistical study might be able to place 
an upper bound on the volume that such unexplored models might occupy 
in the {\it probability space}\/,
this does not translate into any bound on the corresponding volume
that such models might occupy in the {\it model space}\/.
Thus, as long as such models have sufficiently small intrinsic probabilities $p_i$,
their total number can essentially grow  
without bound and yet remain unobservable.

Despite these observations, we are not pessimistic about statistical
explorations of the landscape.
Instead, we feel that efforts to take this exploration seriously 
are important and must continue.
As string phenomenologists, we cannot hope to make progress without ultimately coming to terms
with the landscape.  Given that large numbers of string vacua exist, it is imperative that 
string theorists learn about these vacua and the space of resulting phenomenological possibilities.  As 
already noted in Ref.~\cite{dienes}, the 
first step in any scientific examination of a large data set is that of enumeration and 
classification.  This has been true in branches of science ranging from astrophysics and 
botany to zoology, and it is no different here.
However, before we can undertake this 
monumental enterprise, we will first need to develop an entire toolbox of 
statistical techniques and algorithms 
which are especially constructed for the task at hand.
It is therefore our hope that the methods developed in this paper
will represent one small but useful tool in this toolbox.


\section*{Acknowledgments}
\setcounter{footnote}{0}

This work is supported in part by the National Science Foundation
under Grants PHY/0301998 and PHY/9907949,
by the Department of Energy under 
Grant~DE-FG02-04ER-41298, and by a Research Innovation Award from
Research Corporation. 
KRD wishes to acknowledge
the Galileo Galilei Institute (GGI)
in Florence, Italy, 
and the 
Kavli Institute for Theoretical Physics (KITP) 
at the University of California, Santa Barbara,
for hospitality during the completion of this work.
ML gratefully acknowledges support from the organizers of 
the SUSY 2006 conference and the Santa Fe 2006 Summer
Workshop, where early versions of this work
were presented.


\setcounter{section}{0}   
\Appendix{}
\setcounter{footnote}{0}

Eq.~(\ref{kgolden}) has been derived for the simple case in which 
the population matrix $N_{\alpha i}$ is diagonal.
However, 
 as long as $\alpha$ and $\beta$ are in the same universality class,
it turns out that this result also holds for the more general case
in which our populations $\alpha$ and $\beta$ 
are non-trivially distributed across different probabilities $p_i$.
To see this, let us
begin again with our probability condition
\beq
          P^{(\alpha)}_{\rm new}(D') - P^{(\beta)}_{\rm new}(D'') ~=~ 0~.  
\label{bbal}
\eeq
For convenience, we shall write these expressions in terms of the total number of draws
$D$ rather than the individual counts $d_\alpha$.
In general, these restricted probabilities are given by
\beqn
        P^{(\alpha)}_{\rm new}(D') &\equiv& 
      {1\over\Omega_\alpha} \, \sum_i  \, p_i \left[ N_{\alpha i} - M_{\alpha i}(D')\right]~ 
        \nonumber\\
        P^{(\beta)}_{\rm new}(D'') &\equiv& 
      {1\over\Omega_\beta} \, \sum_i  \, p_i \left[ N_{\beta i} - M_{\beta i}(D'')\right]~ 
\label{restrictedprobs}
\eeqn
where $M_{\alpha i}(D)$ denotes the number of distinct $\alpha$-models already found 
in probability class $p_i$ and  
where $\Omega_{\alpha,\beta}$ are respectively the total probability-space volumes 
of the $\alpha$- and $\beta$-models:
\beq
         \Omega_\alpha~\equiv~ \sum_j p_j N_{\alpha j}~,~~~~~~
         \Omega_\beta~\equiv~ \sum_j p_j N_{\beta j}~.
\eeq
Substituting Eq.~(\ref{restrictedprobs}) into Eq.~(\ref{bbal}) then yields the
condition
\beq
     \sum_i  p_i \left\lbrace
       \left\lbrack {N_{\alpha i}\over\Omega_\alpha} - 
                    {N_{\beta i}\over\Omega_\beta} 
        \right\rbrack ~-~
       \left\lbrack {M_{\alpha i}(D') \over\Omega_\alpha} - 
                    {M_{\beta i}(D'') \over\Omega_\beta} 
        \right\rbrack  \right\rbrace ~=~ 0~.
\label{bigone}
\eeq
Let us now assume that $\alpha$ and $\beta$ are in the same universality class,
as defined through Eq.~(\ref{preuniversalityclass});
the more general definition in Eq.~(\ref{universalityclass}) can be handled
through a reshuffling of indices in what follows.
Given Eq.~(\ref{preuniversalityclass}),
let us define the ratio
\beq
         \gamma ~\equiv ~  {N_{\alpha i}\over N_{\beta i}} ~=~ {N_{\alpha j}\over N_{\beta j}}
             ~~~~~~ {\rm for~all}~~(i,j)~.
\label{eeqiv}
\eeq
We then trivially see that 
\beq
    {N_\alpha \over N_\beta}  ~=~ 
         {\sum_i  N_{\alpha i} \over  \sum_i N_{\beta i}} ~=~ 
         \gamma~
\label{Ns}
\eeq
and
\beq
    {\Omega_\alpha \over \Omega_\beta} ~=~ 
         {\sum_i p_i N_{\alpha i} \over  \sum_i p_i N_{\beta i}} ~=~ 
         \gamma \, {\sum_i p_i N_{\beta i} \over  \sum_i p_i N_{\beta i}} ~=~ \gamma~,
\eeq
whereupon it follows that the term in the first square brackets in Eq.~(\ref{bigone}) vanishes.
Eq.~(\ref{bigone}) thus reduces to the condition 
\beq
         \sum_i p_i 
       \left\lbrack {M_{\alpha i}(D') \over\Omega_\alpha} - 
                    {M_{\beta i}(D'') \over\Omega_\beta} 
        \right\rbrack   ~=~ 0~.
\label{smallcond}
\eeq
However, this condition must hold for {\it all}\/ appropriately balanced pairs $(D',D'')$, since
we want our results to be stable as a function of sample size.
Indeed, there are literally an {\it infinite}\/ number of 
such pairs $(D',D'')$ for which we require that Eq.~(\ref{smallcond}) hold,
leading to a number of distinct constraints (\ref{smallcond}) which is guaranteed to 
exceed the number of probability populations.
(One can prove this last statement through induction.)
Given this, there is only one possible solution:  we must have
\beq
       {M_{\alpha i}(D') \over M_{\beta i}(D'')} ~=~ {\Omega_\alpha\over \Omega_\beta} ~=~\gamma
\eeq
for all $i$.
It then follows that
\beq
          {M_\alpha(D') \over M_\beta(D'') } ~=~ 
        {\sum_i M_{\alpha i}(D')\over \sum_i M_{\beta i}(D'') } ~=~ \gamma~,
\eeq
and in conjunction with Eq.~(\ref{Ns}) this 
yields Eq.~(\ref{kgolden}), as originally claimed. 
Thus, once again, we see that we can overcome all of the biases inherent in the model-generation
process by focusing on the {\it probabilities}\/ for generating new distinct models,
as expressed in Eqs.~(\ref{probbs}) or (\ref{restrictedprobs}),
and by comparing the numbers of models which have emerged at {\it different}\/ points in
the model-generation process at which 
these respective production probabilities are equal.

\vfill\eject
\bibliographystyle{unsrt}

\begin{thebibliography}{99}


\bibitem{landscape}
  S.~Kachru, R.~Kallosh, A.~Linde and S.~P.~Trivedi,
  Phys.\ Rev.\ D {\bf 68}, 046005 (2003)
  [arXiv:hep-th/0301240];\\
  L.~Susskind,
  arXiv:hep-th/0302219.\\
   For popular introductions, see:\\
 R.~Bousso and J.~Polchinski,
  Sci.\ Am.\  {\bf 291}, 60 (2004);\\
   S.~Weinberg,
  arXiv:hep-th/0511037.






\bibitem{abstract}
  M.~R.~Douglas,
  JHEP {\bf 0305}, 046 (2003)
  [arXiv:hep-th/0303194].

\bibitem{abstract2}
  S.~Ashok and M.~R.~Douglas,
  JHEP {\bf 0401}, 060 (2004)
  [arXiv:hep-th/0307049];\\
  F.~Denef and M.~R.~Douglas,
  JHEP {\bf 0405}, 072 (2004)
  [arXiv:hep-th/0404116];\\
  A.~Giryavets, S.~Kachru and P.~K.~Tripathy,
  JHEP {\bf 0408}, 002 (2004)
  [arXiv:hep-th/0404243];\\
   A.~Misra and A.~Nanda,
  Fortsch.\ Phys.\  {\bf 53}, 246 (2005)
  [arXiv:hep-th/0407252];\\
  M.~R.~Douglas,
  Comptes Rendus Physique {\bf 5}, 965 (2004)
  [arXiv:hep-th/0409207];\\
  J.~Kumar and J.~D.~Wells,
  Phys.\ Rev.\ D {\bf 71}, 026009 (2005)
  [arXiv:hep-th/0409218];
  JHEP {\bf 0509}, 067 (2005)
  [arXiv:hep-th/0506252];
  arXiv:hep-th/0604203;\\
  F.~Denef and M.~R.~Douglas,
  JHEP {\bf 0503}, 061 (2005)
  [arXiv:hep-th/0411183];\\
  O.~DeWolfe, A.~Giryavets, S.~Kachru and W.~Taylor,
  JHEP {\bf 0502}, 037 (2005)
  [arXiv:hep-th/0411061];\\
  B.~S.~Acharya, F.~Denef and R.~Valandro,
  JHEP {\bf 0506}, 056 (2005)
  [arXiv:hep-th/0502060];\\
  M.~R.~Douglas and W.~Taylor,
  arXiv:hep-th/0606109;\\
  B.~S.~Acharya and M.~R.~Douglas,
  arXiv:hep-th/0606212;\\
  J.~Shelton, W.~Taylor and B.~Wecht,
  arXiv:hep-th/0607015;\\
   A.~Hebecker and J.~March-Russell,
  arXiv:hep-th/0607120.



\bibitem{susybreakingabstract}
  M.~R.~Douglas,
  arXiv:hep-th/0405279;\\
  M.~Dine, E.~Gorbatov and S.~D.~Thomas,
  arXiv:hep-th/0407043;\\
  M.~Dine, D.~O'Neil and Z.~Sun,
  JHEP {\bf 0507}, 014 (2005)
  [arXiv:hep-th/0501214];\\
  JHEP {\bf 0601}, 162 (2006)
  [arXiv:hep-th/0505202];\\
  M.~Dine and Z.~Sun,
  JHEP {\bf 0601}, 129 (2006)
  [arXiv:hep-th/0506246].



\bibitem{NPcomplete}
  F.~Denef and M.~R.~Douglas,
  arXiv:hep-th/0602072.
 


\bibitem{direct}
  R.~Blumenhagen, F.~Gmeiner, G.~Honecker, D.~Lust and T.~Weigand,
  Nucl.\ Phys.\ B {\bf 713}, 83 (2005)
  [arXiv:hep-th/0411173];\\
  F.~Gmeiner, R.~Blumenhagen, G.~Honecker, D.~Lust and T.~Weigand,
  JHEP {\bf 0601}, 004 (2006)
  [arXiv:hep-th/0510170].\\
  For a review, see:  F.~Gmeiner,
  arXiv:hep-th/0608227.


\bibitem{direct2}
  T.~P.~T.~Dijkstra, L.~R.~Huiszoon and A.~N.~Schellekens,
  Phys.\ Lett.\ B {\bf 609}, 408 (2005)
  [arXiv:hep-th/0403196];
  Nucl.\ Phys.\ B {\bf 710}, 3 (2005)
  [arXiv:hep-th/0411129].



\bibitem{direct3}
  J.~P.~Conlon and F.~Quevedo,
  JHEP {\bf 0410}, 039 (2004)
  [arXiv:hep-th/0409215].
  


\bibitem{dienes}
    K.~R.~Dienes,
  Phys.\ Rev.\ D {\bf 73}, 106010 (2006)
  [arXiv:hep-th/0602286].

   





\bibitem{review1}
  J.~Kumar,
  arXiv:hep-th/0601053.



\bibitem{review2}
   M.~R.~Douglas and S.~Kachru,
  arXiv:hep-th/0610102.




\bibitem{constructions}
   For recent reviews, see:\\
  R.~Blumenhagen, M.~Cvetic, P.~Langacker and G.~Shiu,
  arXiv:hep-th/0502005;\\
   M.~Grana,
  Phys.\ Rept.\  {\bf 423}, 91 (2006)
  [arXiv:hep-th/0509003].







\bibitem{Senechal}
D.~S\'en\'echal,
Phys.\ Rev.\ D {\bf 39}, 3717 (1989);\\
K.~R.~Dienes,
Phys.\ Rev.\ Lett.\  {\bf 65}, 1979 (1990);
Ph.D.~thesis (Cornell University, May 1991), UMI-9131336.


\bibitem{KLT}
   H.~Kawai, D.~C.~Lewellen and S.~H.~H.~Tye,
   Nucl.\ Phys.\ B {\bf 288}, 1 (1987);\\
   I.~Antoniadis, C.~P.~Bachas and C.~Kounnas,
   Nucl.\ Phys.\ B {\bf 289}, 87 (1987);\\
   H.~Kawai, D.~C.~Lewellen, J.~A.~Schwartz and S.~H.~H.~Tye,
   Nucl.\ Phys.\ B {\bf 299}, 431 (1988).




\bibitem{Narain}
   K.~S.~Narain, Phys.\ Lett.\ B {\bf 169}, 41 (1986);\\
   K.~S.~Narain, M.~H.~Sarmadi and E.~Witten, Nucl.\ Phys.\ B {\bf 279}, 369 (1987).


\bibitem{Lerche}
   W.~Lerche, D.~Lust and A.~N.~Schellekens,
   Nucl.\ Phys.\ B {\bf 287}, 477 (1987).




\bibitem{next}
   K.~R. Dienes, M. Lennek, V. Wasnik {\it et al}\/, to appear.




\bibitem{fieldtheory}
  K.~R.~Dienes, E.~Dudas and T.~Gherghetta,
  Phys.\ Rev.\ D {\bf 72}, 026005 (2005)
  [arXiv:hep-th/0412185];\\
  N.~Arkani-Hamed, S.~Dimopoulos and S.~Kachru,
  arXiv:hep-th/0501082;\\
  J.~Distler and U.~Varadarajan,
  arXiv:hep-th/0507090;\\
   B.~Feldstein, L.~J.~Hall and T.~Watari,
  arXiv:hep-ph/0608121.




\end{thebibliography}

\end{document}